\documentclass{article}
\usepackage{amsmath,amssymb,stmaryrd}
\usepackage{amsthm}
\usepackage{authblk}
    \usepackage[english]{babel}
    \usepackage[utf8]{inputenc}
\usepackage{color}

\usepackage{mathtools}
\DeclarePairedDelimiter\ceil{\lceil}{\rceil}
\DeclarePairedDelimiter\abs{\lvert}{\rvert}

\newcommand{\xx}[1]{ [{{\color{red}\bf #1}}] }
\usepackage{algorithm2e}
\usepackage{algpseudocode}
\usepackage{caption}
\usepackage{subcaption}

\DeclarePairedDelimiter\floor{\lfloor}{\rfloor}

\newif\ifsolns
\solnsfalse

\ifsolns
\newcommand{\soln}[1]{\newline \noindent {\bfseries Solution:} {\itshape #1}}
\else
\newcommand{\soln}[1]{}
\fi
\usepackage[margin=2.75cm]{geometry}

\usepackage{fancybox}
\begin{document}
\title{A Quantile-Based Approach to Modelling Recovery Time in Structural Health Monitoring}
\author[1,2]{Alastair Gregory}
\author[1,2]{F. Din-Houn Lau}
\author[1,3]{Liam Butler}
\affil[1]{Lloyd's Register Foundation's Programme for Data-Centric Engineering,
Alan Turing Institute}
\affil[2]{
Department of Mathematics, Imperial College London}
\affil[3]{Cambridge Centre for Smart Infrastructure and Construction, Department of Engineering, University of Cambridge}

\maketitle
\begin{center}
\end{center}

\begin{abstract}

Statistical techniques play a large role in the structural health
monitoring of instrumented infrastructure, such as a railway bridge
constructed with an integrated network of fibre optic sensors. One possible way to reason
about the structural health of such a railway bridge, is to model the
time it takes to recover to a no-load (baseline) state after a train
passes over. Inherently, this recovery time is random and should be
modelled statistically. This paper uses a non-parametric model, based
on empirical quantile approximations, to construct a space-memory
efficient baseline distribution for the streaming data from these
sensors. A fast statistical test is implemented to detect deviations
away from, and recovery back to, this distribution when trains pass
over the bridge, yielding a recovery time. Our method assumes that there are no temporal variations in the
data. A median-based detrending scheme is used to remove the
temporal variations likely due to temperature changes.
This allows for the continuous recording of sensor data with a space-memory constraint.


\end{abstract}
\section{Introduction}

Structural health monitoring (SHM) is used to maintain and monitor
the structural integrity of infrastructure and assets \cite{Farrar}, and has
traditionally been implemented via expensive and time-consuming manual
inspection. Recently, infrastructures have been instrumented with sensors
\cite{Sun, Das, Todd} in an effort to understand more about the
structural health of structures. Typically, the data used in SHM are the vibration responses of the structure. In this work, we use strain data to reason about structural health. It is the hope that data accrued
through sensor networks will improve the maintenance and
inspection of structures. With the sudden
surge in available data sets from these sensors, the development of statistical
methods to complement them must be pursued as a priority \cite{Hernandez}.

A key objective of SHM is to monitor a structure to detect any deterioration in it or an
imminent failure. 
One way to detect long-term deterioration in a structure such as a bridge, is to monitor the time it takes for a structures' dynamic strain response to a specific loading event (e.g. a passing train) to return to some baseline condition (i.e. unloaded state). This duration of time will be referred to herein as the recovery time. It stands to reason that if the structure's recovery time to a specific loading event increases over time, a certain amount of deterioration or damage will have occurred. Where a network of sensors is installed to measure strain at multiple points in a structure, the recovery time at these multiple points can be monitored individually and used to isolate and determine whether deterioration is more prevalent in certain parts of the structure. Therefore, an abundance of data from sensors fitted on to these structures
provides the opportunity to use well-principled statistical techniques to
model such parameters \cite{Noel}. The objective of this paper is to improve the
understanding of this particular aspect of SHM, via a
case-study of railway bridges fitted with a network of discrete fibre optic sensors (FOS). The Cambridge Centre for Smart Infrastructre and Construction (CSIC)
currently installs FOS networks on railway bridges. This work develops methodology with a view to improve the monitoring of railway
bridges supported by data from these sensors, and other instrumented infrastructure in general.

The objective of modelling recovery times of structures from an event, such as
a train passing over a bridge, using data accrued through a sensor network is a
statistical problem. For instance, the recovery times of a bridge are random,
and dependent on various factors i.e. train length, train speed,
environmental factors etc. Statistical methods can be used to model such
recovery times and quantify their uncertainty. This paper builds on some of the statistical techniques
presented in \cite{Lau} for the SHM of railway bridges. Further, we
model the recovery time by understanding the baseline distribution of the FOS strain measurements. Using this baseline distribution, any changes in global behaviour of the bridge can be assessed over time.

The data that are used in this statistical study are considered in a streaming setting
\cite{Lau}, which must be analysed in an online manner. It is assumed
infeasible to store the entire data stream. Therefore, any analysis of the
data in question needs to be space-memory efficient. In addition to this, the
strain data collected from the FOS sensor network arrive at a high frequency (upwards of
250Hz). Many common-place statistical methodologies have been adapted
to fit into this streaming context \cite{Kifer, Tveten, Lall} to include not
only statistical model updates, but inference as well. For example change
detection is an interesting problem in a streaming setting - the goal
being the
automatic detection and a switch of regime given a shift in the underlying
distribution of streamed data. 

To address this, statistical hypothesis tests for changes in distribution have been investigated
for this streaming setting; \cite{Lall} proposes algorithms for the one and two
sample Kolmogorov-Smirnov tests using data that has been streamed. This and
other data streaming literature considers quantile estimation \cite{Greenwald,
  Arandjelovic}. The objective of this estimation is to produce approximate
quantiles of the streaming data, whilst only storing a fixed amount of data
in space-memory. This is required for cases where sensors are continuously
recording. Currently, the FOS-based monitoring system used in this study is only capable of collecting data during short intervals of time (i.e. 1--4 hours). However, the techniques developed in this paper
allow continuous recording from
the sensor network whilst detecting train passage events and estimating the recovery times of them. This is achieved by implementing a space-memory efficient quantile model.

In addition to this, the underlying distribution of the sensor data
considered is difficult to characterise. It is our objective to characterise
this baseline distribution, under no loading, in order to determine deviations
from it and also recoveries to it after a period of loading. This
distribution is not Gaussian, and appears to be bounded. Fitting a
Gaussian to this bounded data, would result in the likelihood of tail
events being over-estimated. Due to the FOS hardware used to collect and pre-process the data, the data is also banded with a discreteness to
the measured values; there are only a very small number of unique
measurements. Our proposed characterisation of the baseline
distribution fits alongside these properties. The non-parametric
approach proposed, improves upon  the work presented in \cite{Lau} that assumes Gaussianity through the use of a linear model.

The data displays temporal variation that is likely due to the sensors
sensitivity to temperature \cite{Kreuzer}. However it is shown in this paper
that the structure of the baseline sensor data distribution remains approximately
the same over time, but has a shifted median due to this temporal
variation. In Sec.~\ref{sec:longtimeseries} a detrending approach based on a moving-median of
the data is utilised to establish a baseline distribution. It will be shown that in sensor data streams that exhibit this temporal variation, the number of unique values increases with the use of detrending. Therefore a quantile-based method of characterising the data is more appropriate in this regime than other such streaming data aggregation methods (e.g. frequency counts \cite{Buragohain, Shrivastava}).

The contribution of this paper is two-fold. First, we introduce a
space-memory efficient and non-parametric quantile model for the data from a
sensor network instrumented on bridges. This model uses the Greenwald-Khanna
algorithm \cite{Greenwald}. Second, we
propose an algorithm that updates the quantile model whilst simultaneously
checking for deviations from the baseline distribution during non-event
periods. During an event, the algorithm sequentially checks for a recovery to
the baseline distribution. This detection is based on a consensus of $p$-values
from each sensor stream, every time a new data point is available. This allows the recovery time of the bridge to be estimated.




The following sections of this paper are now outlined. In the next section, specifics about the FOS data are presented. This is followed by a review of a streaming quantile estimation model used to infer information about this data in Sec.~\ref{sec:quantile_estimation}. A statistical test for quantile outliers, and a corresponding algorithm to detect recovery times from train passing events are presented in Sec.~\ref{sec:statistical_test} and \ref{sec:algorithm} respectively. Results from implementing the algorithm on two test cases are presented in Sec.~\ref{sec:simulations}, whilst results from implementing the algorithm on experimental data are given in Sec.~\ref{sec:FBG_results}.

\section{Case study bridge and sensor data}\label{sec:sensor-data}

\begin{figure}
\centering
\begin{subfigure}{0.5\textwidth}
  \centering
  \includegraphics[width=0.8\textwidth]{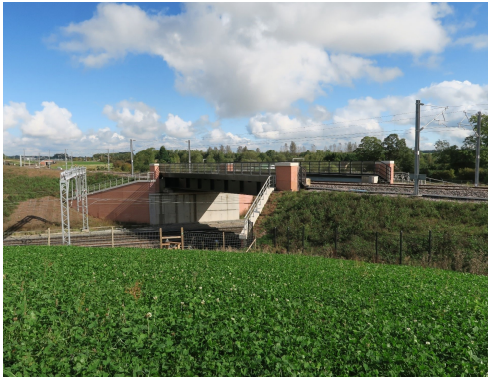}
  \caption{}
  \label{fig:Bridge1}
\end{subfigure}%
\begin{subfigure}{.5\textwidth}
\centering
  \includegraphics[width=0.8\textwidth]{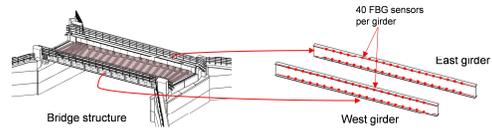}
  \caption{}
  \label{fig:Bridge2}
\end{subfigure}
\caption{An operational railway bridge instrumented with a FOS network (a), and the FOS network topology on the main bridge girders (b).}
\end{figure}

Completed in April 2016, a 26.8 metre half-through steel railway bridge with a concrete composite deck carrying two lines of railway was instrumented with an advanced FOS network during its construction (refer to \cite{Lau} for details). This bridge is shown in Figure \ref{fig:Bridge1}. The sensors are fitted on to the two
girders of the bridge. As such, the strain response along the length of both girders is measured using 20 discrete FOS (spaced at one metre) installed along both the tops and bottom flanges of the girders. Figure \ref{fig:Bridge2} depicts the instrumented railway bridge and the FOS network topology. Together these 80 FOS make up a
network which we aim to utilise in order to infer information about
the structural health of the bridge they are fitted onto. The FOS used as part of this study are based on fibre Bragg gratings (FBGs). Fibre Bragg gratings are periodic variations within the core of a fibre optic cable. Each Bragg grating is created using a specific phase mask which allows the grating to reflect light at a predefined wavelength (i.e. the Bragg wavelength) with all other wavelengths of light passing through. As the fibre optic cable is strained, the Bragg wavelength shifts linearly, allowing the FBG to act as a highly stable and accurate strain sensor. Multiple individual FBGs may be inscribed along a single fibre optic cable (up to 20 FBGs) giving rise to a FBG sensor array. FOS are inherently passive and are non-corrosive and are therefore ideal sensors to be used in permanent long-term structural monitoring systems. Additional details on the operating principles of FBGs are provided in \cite{Kreuzer}.

The data is
measured in wavelength which is typically
converted to strain. This is a useful engineering unit which can subsequently be converted into stress. Denote the wavelength
from sensor $s=1,\ldots,S$ at a time $t$ by $\lambda_{t}^{(s)}$.
Strain measurement is computed by
$$
y^{(s)}_{t} = \frac{(\lambda_{t}^{(s)}-\lambda_{1}^{(s)})}{0.78\lambda_{1}^{(s)}} ,
$$
where $\lambda_{1}^{(s)}$ is the first observed value in the stream of
sensor data. Therefore it is important to note that each stream of
sensor strain data is relative to the first wavelength measurement
observed. Also each strain value has an approximate accuracy of $\pm 4$ microstrain. Throughout this paper we will be working with strain
measurements rather than wavelength. The acquisition rate of the data is 250Hz.

\begin{figure}[t!]
  \centering
  \includegraphics[width=0.75\textwidth]{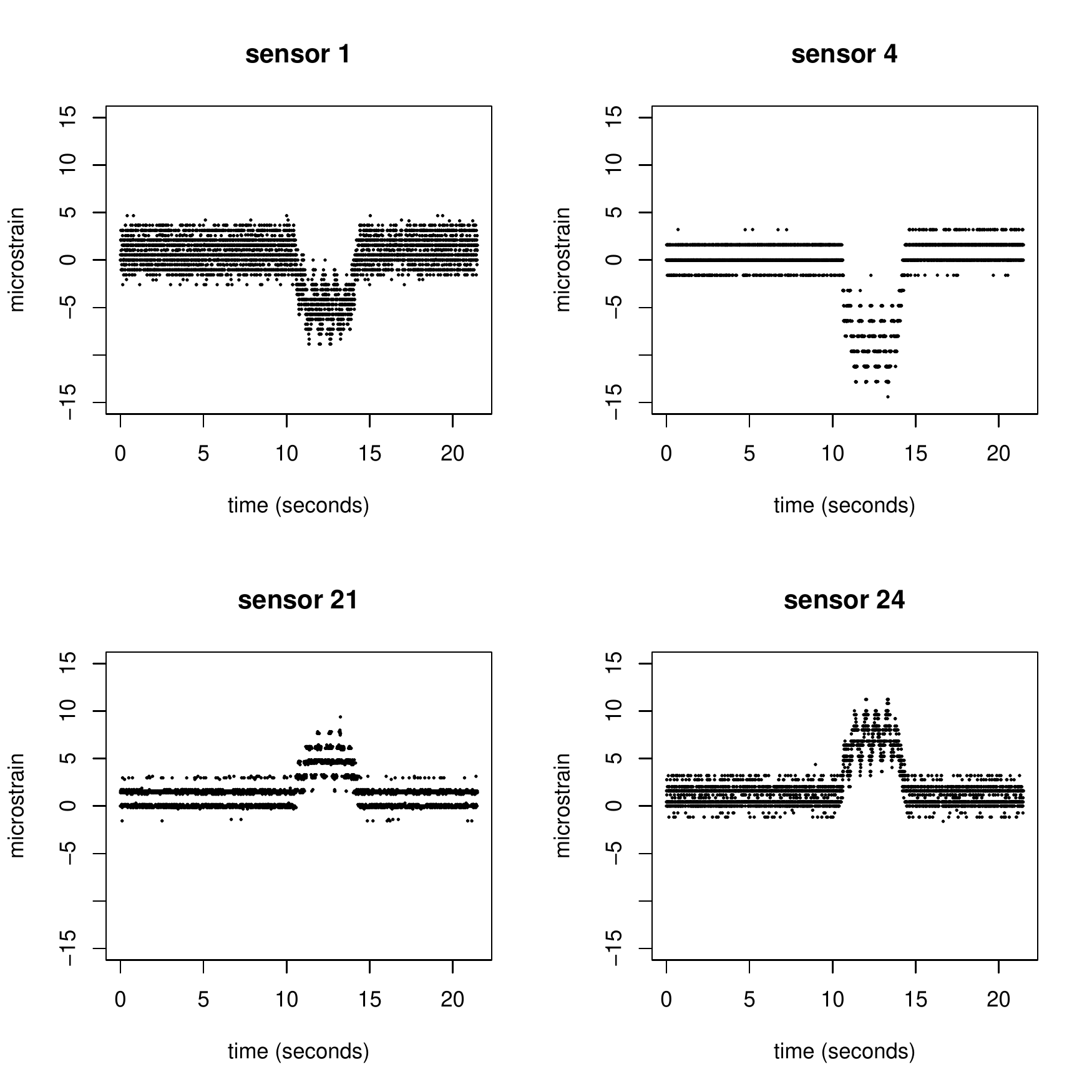}
  \caption{The microstrain measurements from four different FBG
    sensors at the same time during a train passage event.}
  \label{fig:train_sensor_data_1}
\end{figure}

Figure \ref{fig:train_sensor_data_1} presents microstrain measurements
from four sensors, during the same period of time when a train passes
over the bridge. We shall refer to these events as train passage
events. These events are noticeable in the data: a period of large-magnitude strain values. The loading period is the time interval during which the train is actually in contact with the tracks over the bridge and causes these large strain values. A number of preliminary observations can be made from Figure \ref{fig:train_sensor_data_1}, and in particular the \textit{baseline distribution} of the
sensor data. This term will be used throughout the paper to describe
the distribution of the sensor strain measurements,
$\pi^{(s)}(y^{(s)})$, whilst the railway bridge is under no load. The
sensors are placed in different locations on the bridge. As noted above, FBG sensors are located at various locations along the main girders of the bridge. Sensors 1 and
4 are located on the top flange of the east girder, whereas as sensors
21 and 24 are located on the bottom flange of the same girder. Note that
the train passage event feature in the data from top sensors is
inverted for the bottom sensors. This captures the structural mechanics principle of an Euler-Bernoulli beam in sagging bending in which the top flange of the beam is in compression (negative strain) and the bottom flange is in tension (positive strain).

The FOS interrogator hardware and software that were used to collect the FBG data use a set of pre-defined and embedded data processing algorithms written by the manufacturer. As such, the baseline distribution for each sensor exhibits a strong banding feature
where only a few number unique values are observed. Each sensor appears to have a different number of
bands. Also there
appears to be a bound on the values each sensor can take which reflects the strain accuracy of the FBGs (i.e. $\pm$ 4 microstrain).  
Developing a framework where the baseline distributions are
characterised whilst the data are being sequentially observed is an objective of this
paper. 
The work in \cite{Hoshen} considers the problem of characterising
distributions with similar properties, and compares semi-parametric
and non-parametric methods (e.g. using Gaussian mixture models).

\begin{figure}[t!]
  \centering
  \includegraphics[width=0.6\textwidth]{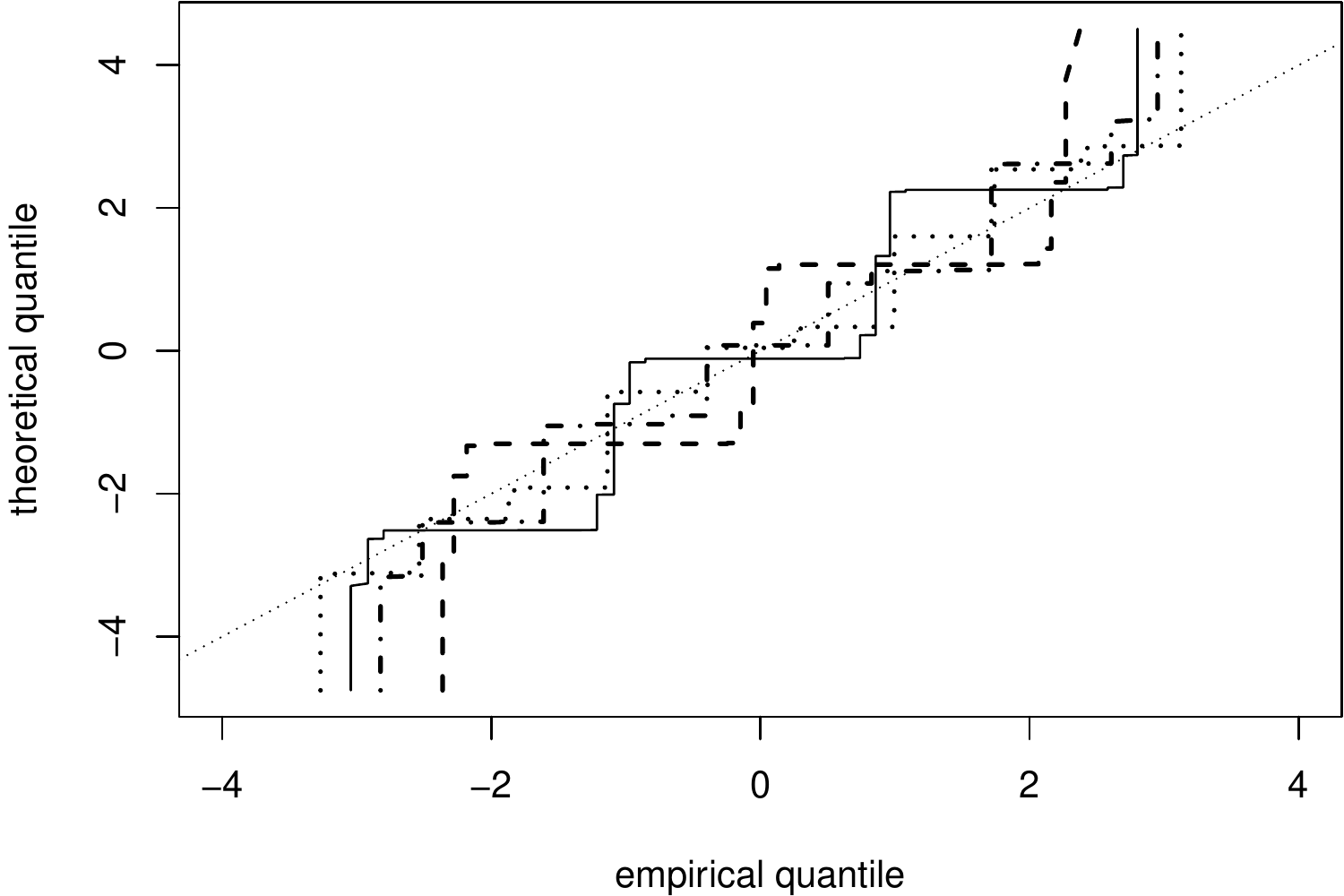}
  \caption{The empirical quantiles of four sensors with microstrain measurements (corresponding to the different line styles), over a period with no train passage events, plotted against the theoretical quantiles of a Gaussian distribution fitted to the mean and variance of each sensors strain data.}
  \label{fig:quantiles_1}
\end{figure}

The boundedness and banding of the sensor distributions suggest that
the data is not Gaussian. The study in \cite{Lau} used a Normal linear
model to predict the sensor data and detect changes from the fitted
model. However the banding of the data suggest that the model used is a simplification. Figure \ref{fig:quantiles_1} shows the empirical quantiles for a set of sensor strain measurements from four different sensors against the theoretical quantiles for a Gaussian distribution fitted to the measurements. These show a disagreement with a Gaussian assumption.

\begin{figure}[t!]
  \centering
  \includegraphics[width=0.6\textwidth]{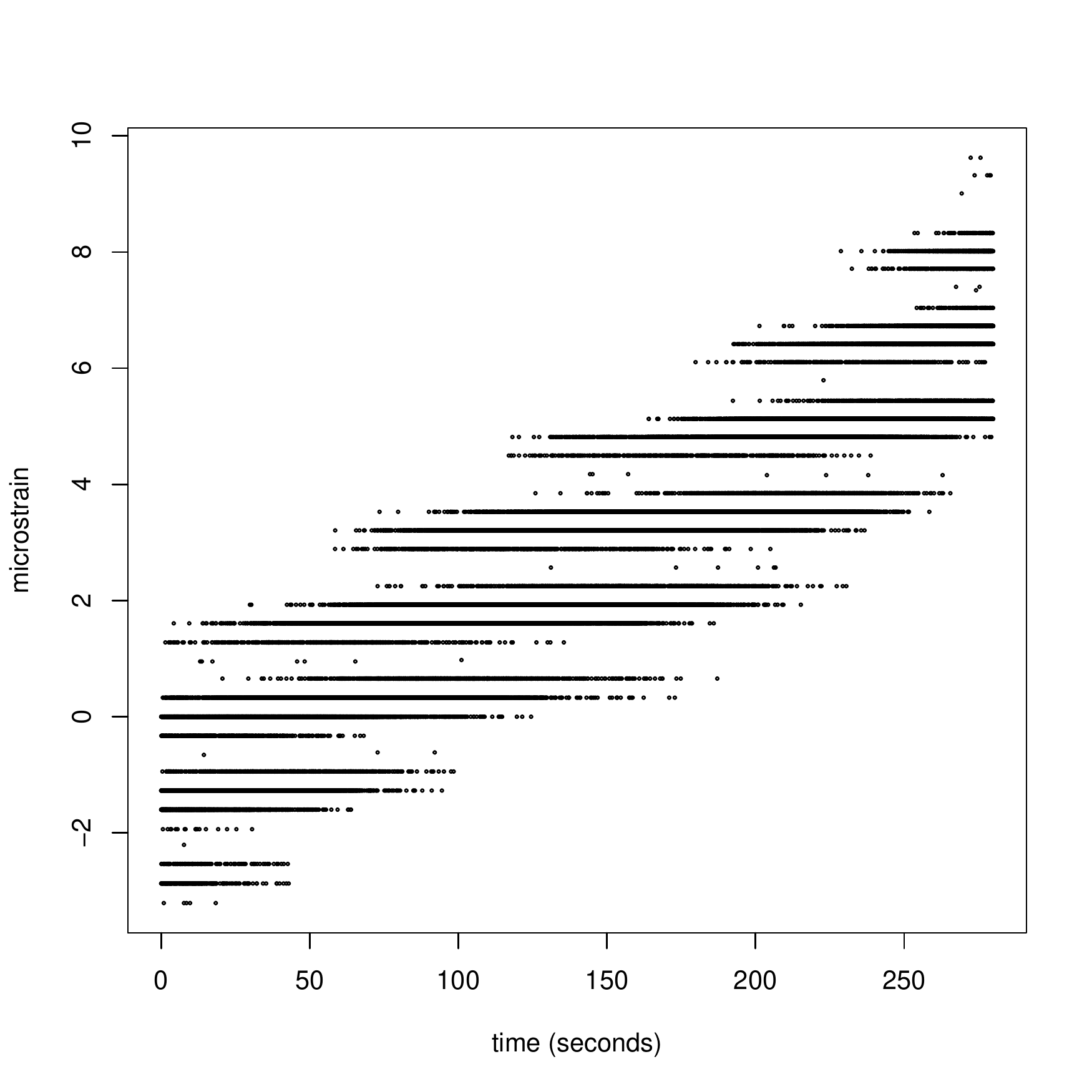}
  \caption{The microstrain measurements from a FBG sensor over a long period of time. Note the temporal shift in the baseline distribution of these measurements, likely due to temperature changes. However the banding structure of the distribution remains throughout the time period.}
  \label{fig:long_sensor_data_1}
\end{figure}

Due to the fact that each strain measurement is dependent on the first
wavelength measurement observed, we expect that every time the FOS
analyser is reset the baseline distribution of the sensor data will shift accordingly. 
The baseline distribution can also shift due to temporal variations,
likely due to temperature. Figure \ref{fig:long_sensor_data_1} shows
the sensor microstrain measurements from one sensor over a longer
period of time in which no train passage event occurs. Note the
temporal variation in strain, however the original banding structure
remains. We would like to detrend the data to obtain a baseline
distribution which is invariant to these temporal shifts. We assume that the deviations away from the baseline sensor
distribution of the detrended sensor data are due to train passage
events only and that the baseline distribution does not change under no
load.

Detrending methods, such as moving-means, can be
used to remove the temporal variation from the sensor data. However,
detrending using moving-means does not keep the original banding
feature of the data \cite{Lau}. This paper proposes to use a moving-median instead. This method also compliments and utilises the
approximate quantile functions described in
Sec. \ref{sec:quantile_estimation}. Using the median maintains the
original banding structure of the sensor strain data. The maintenance
of this banding structure allows us to estimate the baseline
distribution of the data excluding the temporal variations. This is
described in more detail in Sec. \ref{sec:longtimeseries}. Note that some FOS networks may also include temperature sensors alongside strain sensors. The measured temperatures from these sensors may then be utilised to compensate for the effects of temperature on the strain sensors. However, this current study is concerned only with the temporal variations captured by the strain sensors.



\section{Estimating the baseline distribution of sensor data using quantiles}
\label{sec:quantile_estimation}

We model the non-parametric sensor data introduced in the Section \ref{sec:sensor-data} via a quantile function. A quantile function, $F^{-1}(u)$, $u \in [0,1]$, associated with a distribution $\pi(y)$ and the random variable $Y$, returns the $u$-quantile of $\pi_{y}$. The $u$-quantile of a distribution $\pi_{y}(y)$ is the value of $y$ at which the cumulative distribution function, $Pr(Y \leq y)$, crosses $u$.
An empirical quantile function, given a set of data $\big\{y_i\big\}_{i=1}^{n}$, can approximate $F^{-1}(u)$ by
$$
F^{-1}_{n}(u)=\tilde{y}_{\ceil{n \times u}},
$$
where $\tilde{y}_{1}\leq\tilde{y}_{2}\leq\dots\leq\tilde{y}_{n}$ are the order
statistics of $\big\{y_{i}\big\}_{i=1}^{n}$ and $\ceil{\cdot}$ denotes the ceiling function.
In the case of absolutely continuous $y$, this is a consistent estimator of $F^{-1}(u)$. 
In the case of streaming data, where it is not feasible to store (and
continuously sort) the whole sequence of data
$\big\{y_{i}\big\}_{i=1}^{n}$, one can estimate these empirical
quantiles. The studies \cite{Greenwald, Arandjelovic} cover this
problem for example. We will use the Greenwald and Khanna (GK) algorithm \cite{Greenwald}. This method is efficient in the sense that it accurately estimates the quantile function with
significantly less than $n$ data points. The algorithm is described briefly in the next section.

The GK algorithm returns a \textit{quantile
  summary}, $Q_{n}$, containing $L\in\mathbb{Z}^{+}$ stored tuples of
length 3. Each tuple contains one of the $n$ values that has been seen
in the data so far. The method is space-memory efficient since $L \ll
n$. The input to this algorithm is a stream of data, and a single accuracy parameter $\epsilon$. One can query the summary to return an approximation, $Q_{n}(u)$ to $F^{-1}_{n}(u)$, for a $u \in [0,1]$. The summary guarantees
$$
Q_{n}(u) \in [F^{-1}_{n}(u-\epsilon),F^{-1}_{n}(u+\epsilon)].
$$
The next section explains how the algorithm works.



\subsection{The Greenwald-Khanna algorithm}

The GK algorithm works by keeping a set of tuples
$(v_{i},g_{i},\Delta_{i})$, for $i=1,\ldots,L$. Let the quantile
summary be given by
$Q_{n}=\big\{(v_{1},g_{1},\Delta_{1}),\ldots,(v_{L},g_{L},\Delta_{L})\big\}$. The
$v_{i}$'s are sorted so that $v_{1}\leq v_{2} \leq ...\leq v_{L}$. These represent a
subset of the values in the stream, that have been retained in the
summary. Each retained $v_{i}$ value is selected to cover a region of values in the original stream that have ranks between $r_{min}(v_{i})$ and $r_{max}(v_{i})$ in the original stream. The values $g_{i}$ and $\Delta_{i}$ in each tuple contain the information needed to infer these minimum and maximum ranks. The $g_{i}$'s represent $r_{min}(v_{i})-r_{min}(v_{i-1})$, and the $\Delta_{i}$'s represent $r_{max}(v_{i})-r_{min}(v_{i})$ (the range of ranks that $v_{i}$ cover the original stream). At any time, one can compute the maximum and minimum ranks for each summary element $v_{i}$, for $i=1,\ldots,L$, via
$$
r_{min}(v_{i}) = \sum^{i}_{j=1}g_{j}
$$
and
$$
r_{max}(v_{i})=\sum^{i}_{j=1}g_{j}+\Delta_{i} .
$$
To find the values from the stream to retain in the summary, and to obtain the $L$ tuples $(v_{i},g_{i},\Delta_{i})$, the algorithm iterates over two functions, \textbf{INSERT} and
\textbf{COMBINE}, as follows:

\begin{itemize}
\item[]\textbf{COMBINE}: When $$\left(t \mod \left(
      \floor*{1/(2\epsilon)} \right) \right)= 0,$$ one can combine
  multiple adjacent tuples,
  $$\left(v_{i-k},g_{i-k},\Delta_{i-k}\right),\dots,\left(v_{i},g_{i},\Delta_{i}\right),$$
 into one if $g_{i-k}+\ldots +g_{i}+\Delta_{i}\leq 2\epsilon t$. Note that the smallest value must in the summary must be maintained (therefore maintaining the smallest value in the stream), by requiring that $(i-k)>1$ in the explanation of combine above. As $v_{i}$ is kept in each iteration of combine, the largest value in the summary is also maintained, therefore maintaing the largest value in the stream.

\item[] \textbf{INSERT}: Insert the tuple, $(v,1,\Delta)$, where $v$ is the new value in the stream at time $t$, between the tuples $(v_{i},g_{i},\Delta_{i})$ and $(v_{i+1},g_{i+1},\Delta_{i+1})$ where $i$ satisfies $v_{i}\leq v<v_{i+1}$. Here, $\Delta=g_{i}+\Delta_{i}-1$. If $v$ is larger or smaller than every $v_{j}$, for $j=1,...,L$, then insert $(v,1,0)$ at the end or start of the summary.
\end{itemize}

\begin{figure}[t!]
  \centering
  \includegraphics[width=0.6\textwidth]{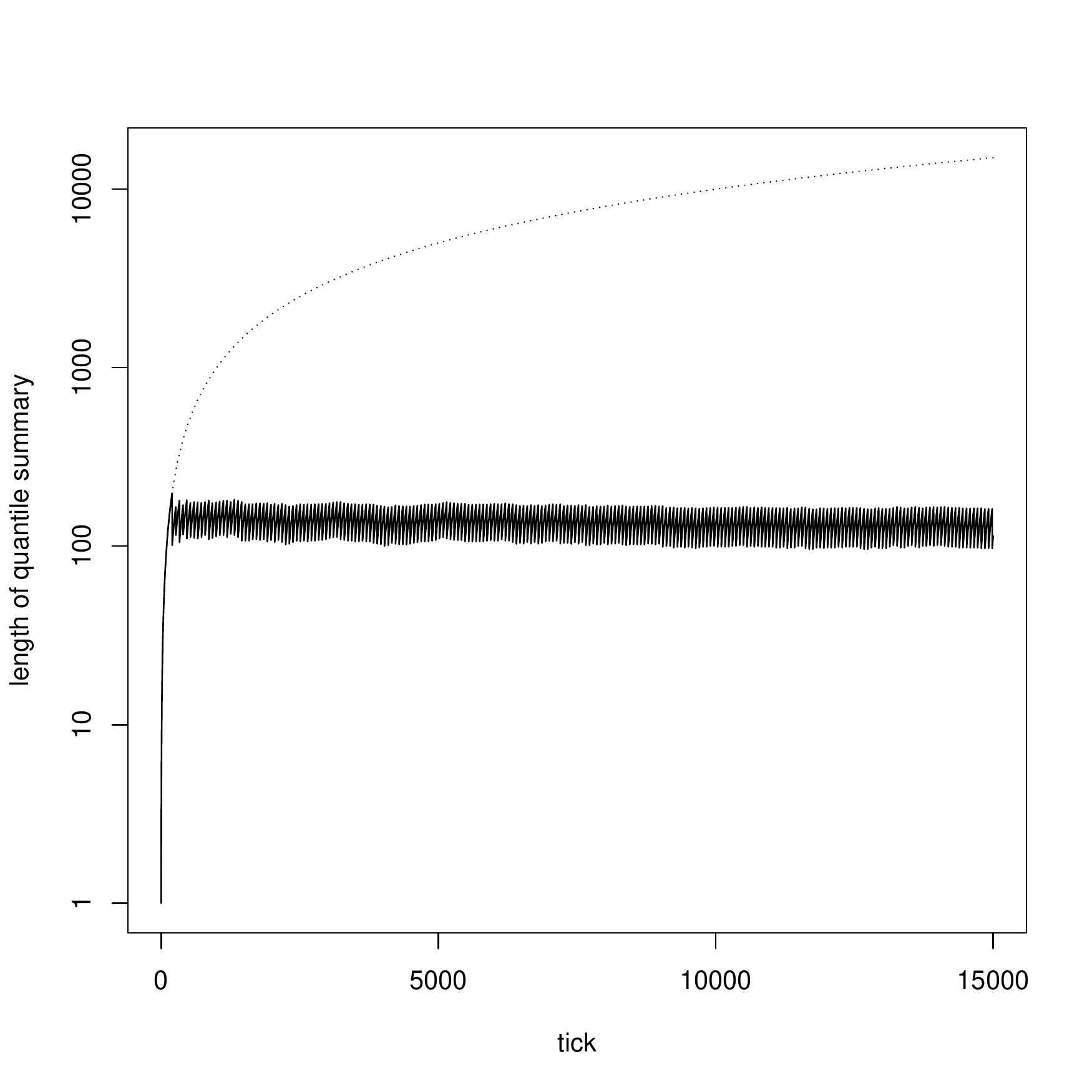}
  \caption{Number of elements in a quantile summary (solid) with $\epsilon=0.0075$, taken over a stream of values sampled from the standard normal distribution. As the number of elements in the stream (dotted) increases, the quantile summary length remains approximately constant.}
  \label{fig:quantile_summary_backgroundsensor}
\end{figure}

The algorithm is fast, and can run online indefinitely with fixed
space-memory. For example, Figure
\ref{fig:quantile_summary_backgroundsensor} shows the number of
elements in a quantile summary (with $\epsilon=0.0075$) taken over a
stream of values sampled from the standard normal
distribution. It updates the quantile summary with this stream of values at a speed of $\sim$700Hz on a midrange HP laptop. For more details about the GK algorithm see \cite{Greenwald}. Many studies have used this algorithm to obtain a summary of quantiles in a streaming data setting. Such an example is \cite{Lall} , where a streaming Kologorov-Smirnov hypothesis test uses a quantile summary constructed in this way. The next section uses a similar approach to \cite{Lall} in constructing a statistical test using a quantile summary, so that a detection algorithm for recovery times from train passage events can be established in Sec. \ref{sec:detection}.

\subsection{Inverse quantile summary queries}

It is proposed in \cite{Lall} that the quantile summaries described
earlier in this section can approximate the empirical distribution
function for a data stream. 
In other words, one can inversely query a quantile summary
$Q_{n}$. This is achieved by a binary search of the summary values $v_{i}$, choosing the index $j$ for which $y$ satisfies
\begin{equation}
j=
\begin{cases}
\min(i; y\geq v_{i}),& y\geq v_{1}\\
0,& y<v_{1},
\end{cases}
\end{equation}
where $v_0=-\infty$. Then one takes
\begin{equation}
\begin{cases}
Q_{n}^{-1}(y)=r_{max}(v_{j}), & j>0\\
Q_{n}^{-1}(y)=0, & j=0.
\end{cases}
\end{equation}
This returns an approximation to the empirical distribution function
$F_{n}(y)=\sum^{n}_{i=1}\textbf{1}_{(y\geq y_{i})}$ with an accuracy
of $\pm 3\epsilon$ \cite{Lall} where $\textbf{1}$ is the indicator
function.

The form of the statistical test used in the next section relies on being able to estimate $p$-values, by evaluating the empirical distribution function of the stream (or more specifically the summary approximation of it). Therefore we wish to have smoothed tails for our empirical estimates, as having zero $p$-values for any $y \geq \tilde{y}_{n}$ or $y<\tilde{y}_{1}$ overestimates the probability of accepting an experimental hypothesis in the test (by always accepting it). Smoothing in the tails is implemented by
\begin{equation}
\tilde{Q}_{n}^{-1}(y) =
\begin{cases}
\frac{2}{n}\phi_{h}(y-v_{1}), & y<v_{1}\\
Q^{-1}_{n}(y), & y \in [v_{1},v_{L}]\\
1-\frac{2}{n}\phi_{h}(v_{L}-y), & y > v_{L}.
\end{cases}
\end{equation}
where $\phi_{h}(y)$ is the distribution function of $N(0,h)$. As $h\to 0$, we have $\tilde{Q}_{n}^{-1}(y) \to Q_{n}^{-1}(y)$ in probability for all $y \in \mathbb{R}$. These kernels are used in well known smoothing distribution function estimates within \cite{Cheng}, however here we only incorporate them at the tails of the empirical distribution. An important property of the Greenwald-Khanna algorithm is that the smallest and largest value in the stream are kept in the summary at all times. Denote these values $\tilde{y}_{1}$ and $\tilde{y}_{n}$. Therefore note that $Q_{n}^{-1}(y)=F_{n}(y)$ when $y \leq \tilde{y}_{1}$ or $y \geq \tilde{y}_{n}$. Using this, one notes that in the tails, $y \notin [\tilde{y}_{1},\tilde{y}_{n})$, we have an error of
$$
\abs{F_{n}(y)-\tilde{Q}_{n}^{-1}(y)} = \abs{Q^{-1}_{n}(y)-\tilde{Q}^{-1}_{n}(y)} \leq 1/n.
$$
Given that the quantile summary is not space-memory efficient if $\epsilon \leq 1/n$, then this error should not exceed $\pm 3 \epsilon$ from the approximation $Q^{-1}_{n}(y)$, for $y \in [\tilde{y}_{1}, \tilde{y}_{n})$. 


\section{Estimating recovery times from train passage events}

The following two subsections explore how the recovery time of an instrumented bridge following a train passage event may be estimated, using a baseline sensor distribution constructed using the techniques introduced in the previous section.

\subsection{A two-threshold statistical test for anomalies}
\label{sec:statistical_test}

In this subsection, we propose a two-threshold statistical test to detect deviations away
from the baseline sensor distribution, given by the quantile model
introduced in Sec. \ref{sec:quantile_estimation}. This statistical
test is based on the anomaly detection method used in \cite{Lau}, and
tests the null hypothesis that strain data from all the sensors fitted onto the bridge are in the baseline state.
To implement the test, we first compute $p$-values of each new sensor data point with respect to the baseline sensor distributions using a smoothed inverse query of the quantile models. As each new sensor point could be a deviation from the baseline distribution at either tail, the $p$-values take the form of
\begin{equation}
p^{(s)}_{t}=\min\left(1-\tilde{Q}^{-1}_{t-1}(y^{(s)}_t),\tilde{Q}^{-1}_{t-1}(y^{(s)}_t)\right).
\label{equation:p-value}
\end{equation}
These evaluate the potential of each new data point of being extreme
in comparison to the baseline sensor distribution. This statistic requires just one inverse query of the quantile summary for each sensor, and can therefore be carried out efficiently after each new set of sensor data points becomes available. We then combine $p$-values from all the sensors using Fishers method \cite{fisher1925statistical}, to collect a consensus of whether the sensor data from the bridge as a whole has deviated from it's baseline state. The $\chi^2$ test statistic is given by
\begin{equation}\label{eq:chisq_stat}
    X^2 =  -2 \sum_{s=1}^S \log(p_t^{(s)}).
  \end{equation}
  Under the null hypothesis the statistic $(X^{2}+c)$, where $c=2S\log(0.5)$, follows a $\chi^2_{2S}$ distribution. To see this, consider sampling $q_i \sim U[0,1]$ i.i.d. for $i=1,...S$. These represent $S$ $p$-values under the null hypothesis, for a continuous underlying distribution. Then $-2\sum^{S}_{i=1}\log(q_{i})$ follows a $\chi^2_{2S}$ distribution. Let $p_i=\min(1-q_i,q_i)$, as in (\ref{equation:p-value}), then $p_i \sim U[0,0.5]$. Therefore the statistic,
  $$
  \left(-2\sum^{S}_{i=1}\log(p_{i})\right)+c=-2\sum^{S}_{i=1}\log(q_{i}),
  $$
  follows a $\chi^2_{2S}$ distribution.
  
  We
  use two thresholds, $k_{l}$ and $k_{u}$, where $k_l\leq k_u$. When a data stream is deemed in the baseline state and the value of
$X^2$ exceeds $k_u$, a train passage event is signalled. On the other
hand, when a data stream is signalling a train passage event
and the $X^2$ value drops below $k_l$, a return to the baseline
distribution is signalled. Formally, the threshold $k_{u}$ corresponds to the confidence in the null
  hypothesis being rejected when the sensor data is currently in the
  baseline state. The threshold $k_{l}$ corresponds to the confidence
  in the null hypothesis not being rejected when the train passge
  event is currently occurring. The value $k_{l}$ should be much less
  than $k_{u}$.


The work in \cite{Lall} considers using the Kolmogorov-Smirnov tests on the quantile summary to construct a similar type of statistical test for divergence from a distribution. However we would like to obtain a test for the deviation from (and recovery to) the baseline distribution after every new data point. Since the train passage events only last for a few seconds, we require a high temporal resolution in recovery time estimates. Whilst this can be done by using the methodology in \cite{Lall}, it would be inefficient as one needs to evaluate the approximate empirical distribution function at each of the $v_{i}$'s in the quantile summary every time the test is implemented. A drawback of implementing the proposed test after every new data point is that false discoveries / deviations from the baseline distribution can occur more frequently. A future study could concentrate on the use of non-restarting and controllable CUSUM charts \cite{Gandy} to reduce this risk.

\subsection{Algorithm}
\label{sec:algorithm}
\label{sec:detection}

This subsection introduces an algorithm (Algorithm \ref{algo:algo1}) that
combines the quantile model of the baseline distribution of the sensor
data (Sec. \ref{sec:quantile_estimation}) and the statistical test for anomalies (Sec. \ref{sec:statistical_test}), to estimate the recovery time from train passage events. The algorithm recursively tests the current sensor data point against the baseline sensor distribution, and then updates the quantile model of the baseline sensor distribution with this data point if no anomaly is detected.

\begin{algorithm}
\caption{Recovery times from train passage events}\label{algo:algo1}
  \KwData{Strain data stream $y_t^{(s)}$ for $s=1,\dots, N_S$ sensors and
    $t=1,2,\dots$}

  \KwIn{Threshold values: $k_u>0$, $k_l>0$\; Minimum time to test for anomalies: $\tau$.}

  \KwOut{Estimated sets of start times $T_S$ and end times $T_E$ }

  Initialise $C=0$\;


\For{$t=1,2,\dots$}{

\If{$t>\tau$}{
Compute $p$-value for each sensor using $y^{(s)}_{t}$\;
Compute the $\chi^2$ test statistic, $X^2$ (see Sec. \ref{sec:statistical_test} for details) \;
}

\If{$X^2 > k_u$ and $C=0$}{
  Set $C=1$\;
  Append $t$ to the set $T_S$\;
}

\If{$X^2 < k_l$ and $C=1$}{
 Set $C=0$\;
 Append $t$ to the set $T_E$\;
}

\While{$C=0$}{ Update quantile summaries $Q_{t}^{(s)}$ using
  $y^{(s)}_{t}$, for $s=1,\dots,S$ (see Sec. \ref{sec:quantile_estimation} for details)\; }
}

\end{algorithm}

The recovery time of the sensor network is estimated by $T_{E}(k)-T_{S}(k)$
for each event $k=1,2,3,\ldots$. 
With the exception of the $\chi^2$ test, the algorithm is run
separately for each sensor, and therefore can be run in parallel. This
is an improvement to the predictive linear model proposed in
\cite{Lau}, which used information from all other sensors for each sensor.

\section{Simulations}
\label{sec:simulations}

In this section, we implement Algorithm \ref{algo:algo1} on two test
cases. Both test cases feature data,
$\big\{y_{t}\big\}_{t=1,2,3,...}$, sampled from a baseline
distribution $f_{1}(y)$, that at a certain time $t=T_{1}$ instantly
changes to being sampled from another distribution $f_{2}(y)$. The return
to the baseline distribution is then either gradual or instantaneous
depending on the test case. These cases are designed to replicate a
single train passage event across a sensor network. The distributions
$f_1$ and $f_2$ are assumed to be unknown in our algorithm. The
distribution $f_{1}$ will have a multimodal structure, similar to
that found in the sensor baseline distribution (see Figure \ref{fig:train_sensor_data_1}). Set the start time $T_1=1000$, the end time $T_{2}=2000$ and
total number of datapoints $T_{3}=3000$. The actual recovery time of these simulations is given by the difference between end and start times, $T_{2}-T_{1}$. Define the mixture distribution $g^{(t)}(y)$ at each time $t \in [1,T_3]$ by
$$
g^{(t)}(y)=w^{(t)}f_1(y)+(1-w^{(t)})f_2(y),
$$
where in case (1),
\begin{equation*}
w^{(t)}=
\begin{cases}
1, & t< T_{1},\\
0, & T_{1}\leq t\leq T_{2},\\
1, & t>T_{2},
\end{cases}
\end{equation*}
and in case (2),
\begin{equation*}
w^{(t)}=
\begin{cases}
1, & t< T_{1},\\
0, & T_{1}\leq t\leq (T_{2}+T_{1})/2,\\
\frac{\left(2t-(T_{2}+T_{1})\right)}{(T_{2}-T_{1})}, & (T_{2}+T_{1})/2<t \leq T_{2},\\
1 & t>T_{2}.
\end{cases}
\end{equation*}

\begin{figure}[t!]
  \centering
  \includegraphics[width=0.7\textwidth]{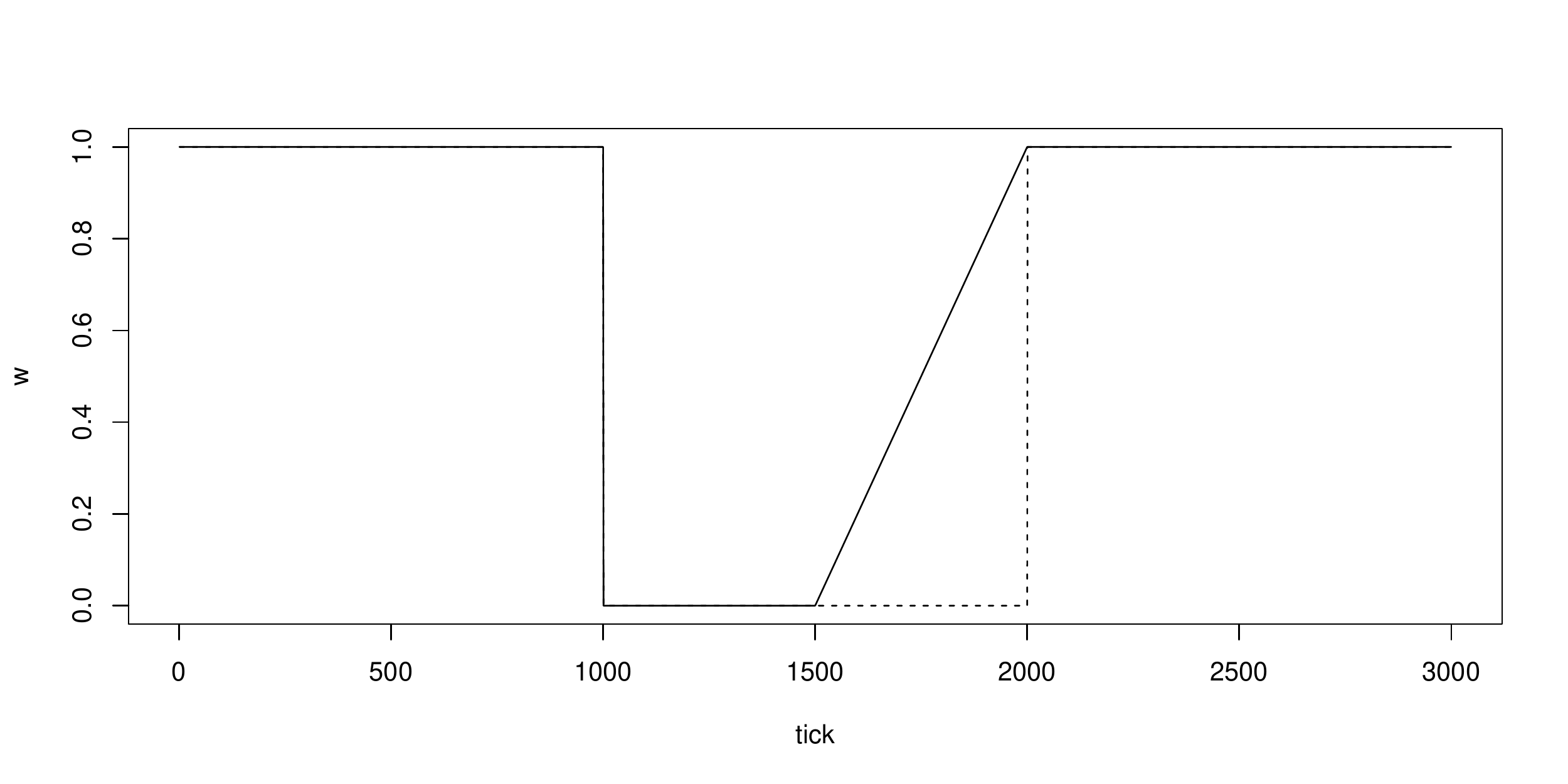}
  \caption{The values of $w^{(t)}$ for test case (1) (dashed) and test
    case (2) (solid).}
  \label{fig:simulation_setup}
\end{figure}

\begin{figure}
\centering
\begin{subfigure}{0.5\textwidth}
  \centering
  \includegraphics[width=0.8\textwidth]{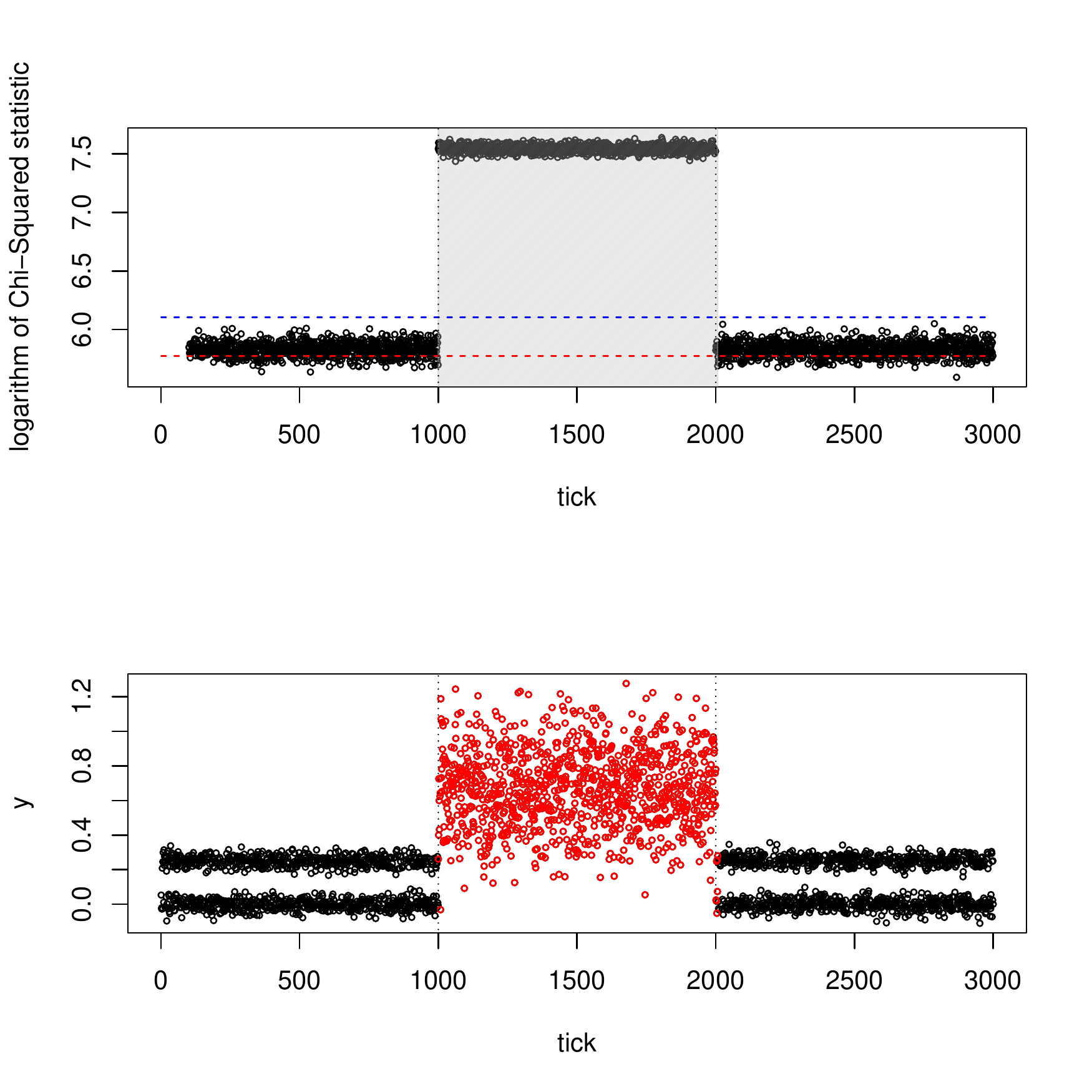}
  \caption{}
  \label{fig:test_case_gradual_change_recoveryalgorithma}
\end{subfigure}%
\begin{subfigure}{.5\textwidth}
\centering
  \includegraphics[width=0.8\textwidth]{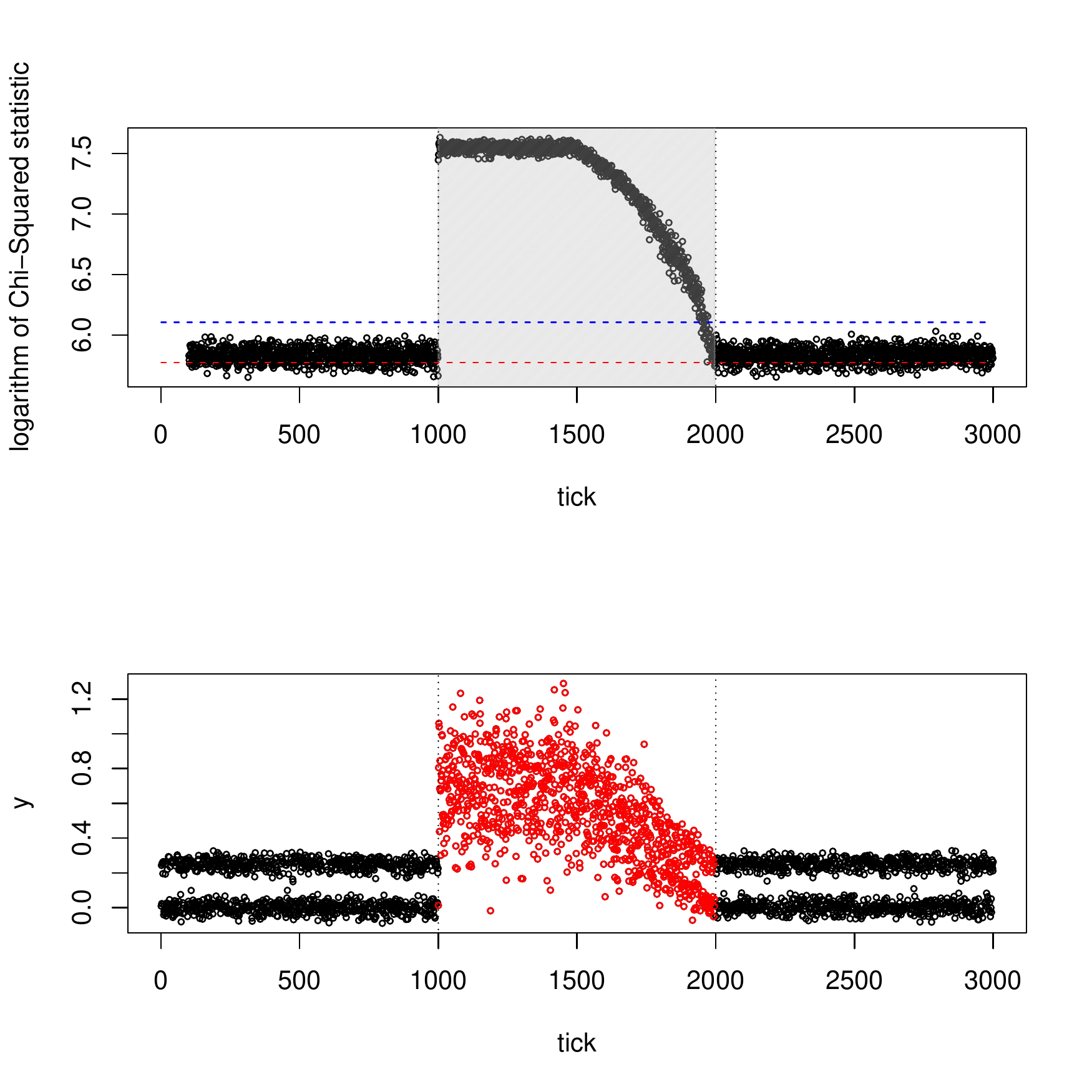}
  \caption{}
  \label{fig:test_case_gradual_change_recoveryalgorithmb}
\end{subfigure}
\caption{Plots of $\log(X^2)$ (top panels) and the data $\big\{y^{(1)}_{t}\big\}_{t=1}^{T_{3}}$ (bottom panels) for the first (a) and second (b) test cases. Black dashed lines show the prescribed start and end points, $T_{1}$ and $T_{2}$, and the red data points / grey rectangle show the detected recovery period. In the top plots, blue and red dashed lines show the values of $\log(k_{u})$ and $\log(k_{l})$ respectively.}
\end{figure}

The values of $w^{(t)}$ for the two cases are shown in Figure \ref{fig:simulation_setup}. In these simulations, the true event start time $T_1$ and
end time $T_2$
are known. Algorithm \ref{algo:algo1} outputs an estimate of the start
and end times, $T_{S}$ and $T_{E}$, and therefore the recovery time. We gauge the performance of our algorithm by comparing
these times. The parameters utilised in Algorithm \ref{algo:algo1} are chosen to be $\epsilon = 0.0075$, $k_{u}=448.5$, $k_{l}=321.6$, $h=0.2$ and $\tau=100$.
For each case, we sample $S=100$ independent data streams,
$\big\{y^{(s)}_{t}\big\}_{t=1}^{T_{3}}$, for $s=1,...,S$, to simulate
multiple sensors. To obtain an empirical distribution of estimated recovery
times ($T_{E}-T_{S}$), 1000 iterations of the simulations are
computed. The plots of
$\big\{y^{(1)}_{t}\big\}_{t=1}^{T_{3}}$ and $\log(X^2)$ for a single iteration of the first and second test cases are shown in
Figures \ref{fig:test_case_gradual_change_recoveryalgorithma} and \ref{fig:test_case_gradual_change_recoveryalgorithmb} respectively .

\begin{figure}
\centering
\begin{subfigure}{0.5\textwidth}
  \centering
  \includegraphics[width=0.8\textwidth]{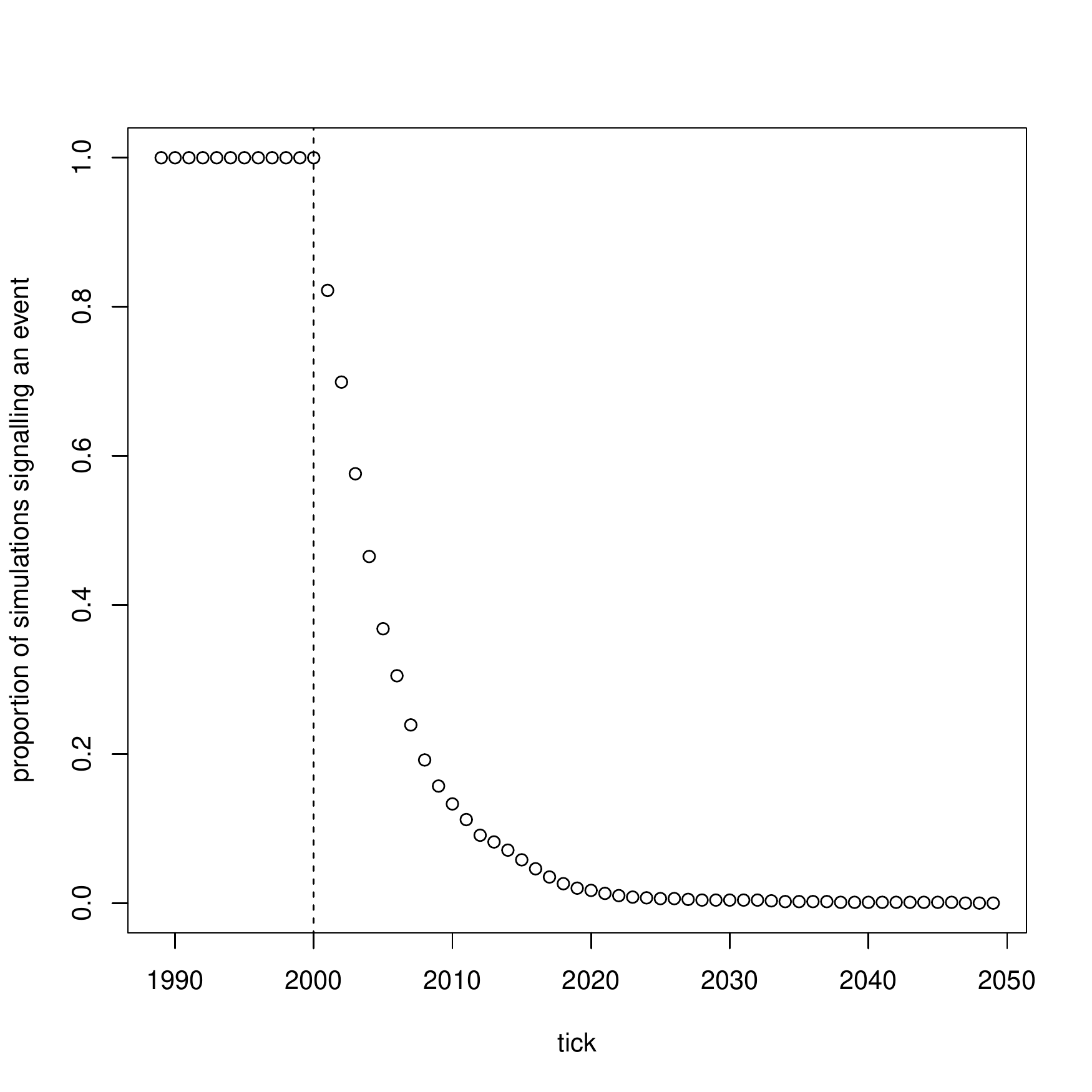}
  \caption{}
  \label{fig:cdf_of_simulations_recoverytime_testcase1}
\end{subfigure}%
\begin{subfigure}{.5\textwidth}
\centering
  \includegraphics[width=0.8\textwidth]{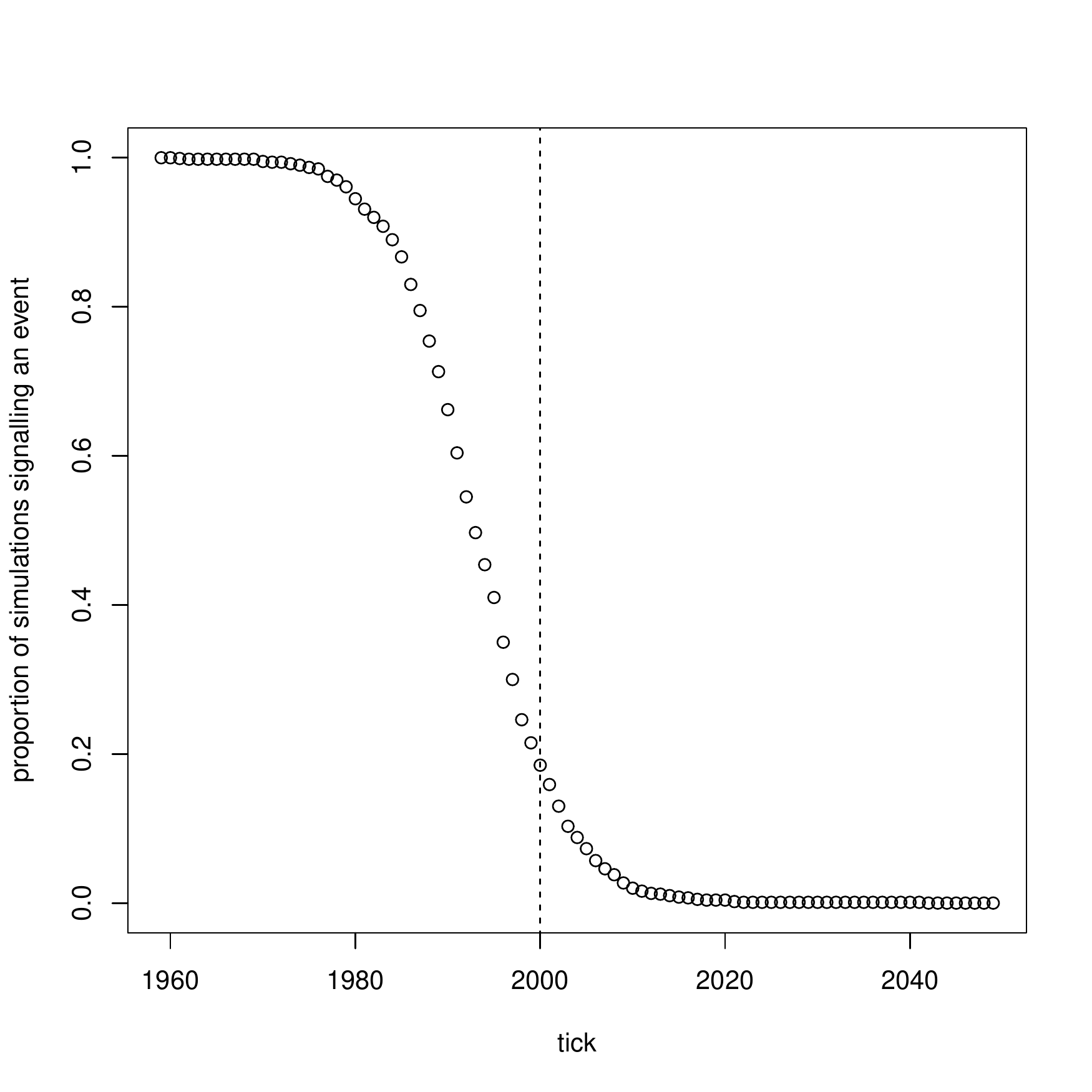}
  \caption{}
  \label{fig:cdf_of_simulations_recoverytime_testcase2}
\end{subfigure}
\caption{Empirical distribution function of $T_{E}$
  for all 1000 iterations of the first (a) and second (b) test cases. The black dashed line represents $T_{2}$.}
\end{figure}

For both of the test cases, the immediacy of the deviation from
$f_{1}$ to $f_{2}$ at $T_{1}$ results in all 1000 iterations
obtaining the exact value of $T_{S}=T_{1}$. Here $k_{l}\ll k_{u}$ to prevent false recoveries back to $f_{1}$. To
evaluate the estimated event end times, and thus the recovery times
for both test cases, the empirical distribution functions of $T_{E}$ are
shown in Figures \ref{fig:cdf_of_simulations_recoverytime_testcase1} and \ref{fig:cdf_of_simulations_recoverytime_testcase2} respectively. For the first test
case, none of the 1000 iterations estimate the recovery before $(T_{2}+1)$, the first point since $T_{1}$ that the data is not sampled from $f_{2}$. For the second test case, there is a mixture of recoveries before and after $T_{2}$ amongst all of the iterations. Manually or adaptively tuning $k_l$, the critical value corresponding to our confidence in the data returning to the baseline distribution, could be implemented here to obtain estimates of $T_{E}$ closer to $T_{2}$.



\section{Application to FBG sensor data}
\label{sec:FBG_results}

We now use Algorithm \ref{algo:algo1} to estimate the recovery times
of the bridge installed with FBG sensors. The data were
recorded in July 2016 and May 2017, where 11 train passage events
occurred in 2016 and 24 in 2017. Algorithm \ref{algo:algo1} is used to
estimate the recovery times for all train passage events and construct
baseline quantile functions for each sensor. The parameters utilised
in Algorithm \ref{algo:algo1} are chosen to be $\epsilon=0.0075$,
$k_{u}=464.7$, $k_{l}=238.9$, $h=0.3$ and $\tau=500$. 



\begin{figure}[t!]
  \centering
  \includegraphics[width=0.6\textwidth]{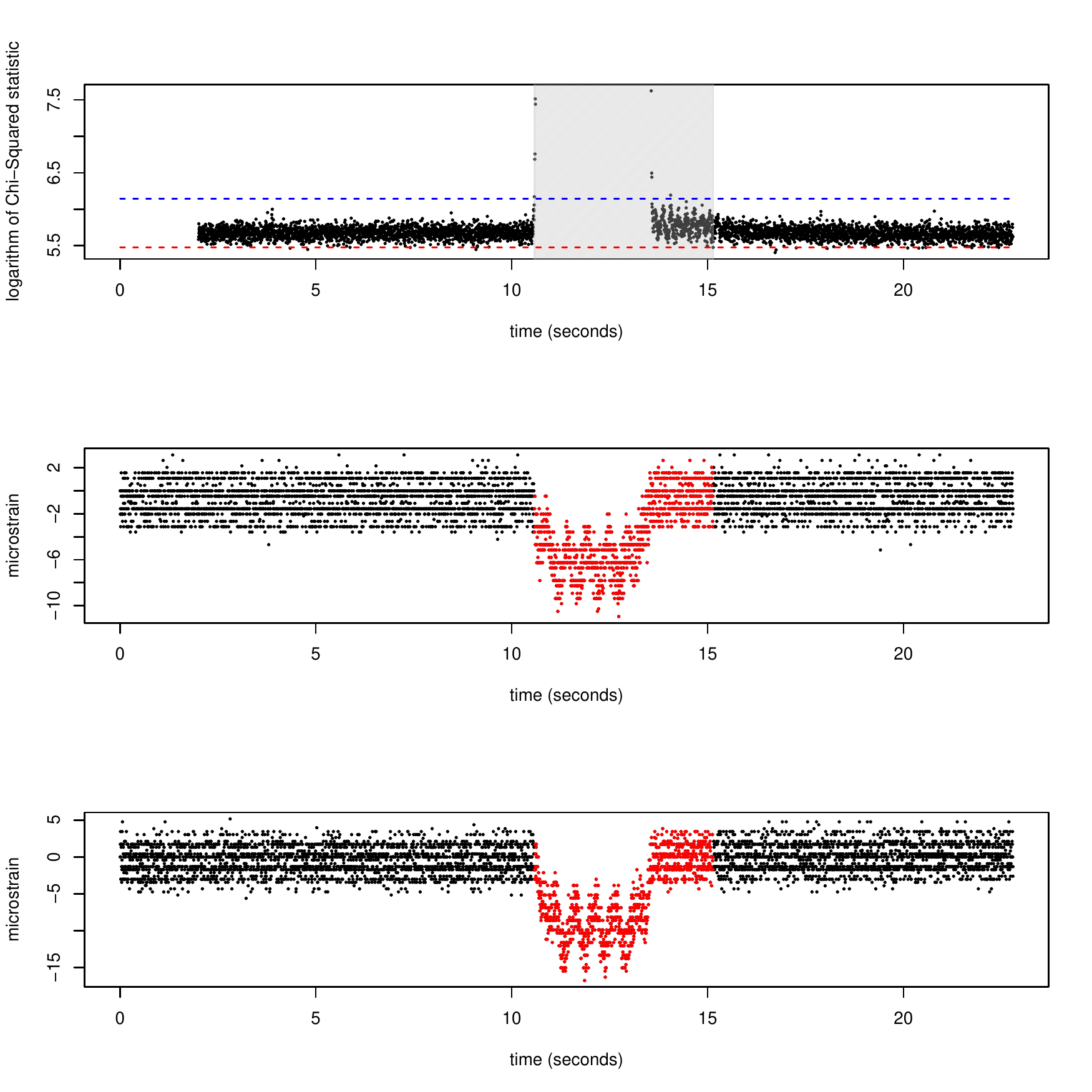}
  \caption{Plot of $\log(X^2)$ (top panel),  the data from sensor 1
    (middle panel) and the data from sensor 20 (bottom panel) that includes a single train passage event. The red data points / grey rectangle show the detected recovery period. In the top plot, blue and red dashed lines show the values of $\log(k_{u})$ and $\log(k_{l})$ respectively.}
  \label{fig:sensor_case_log_recoveryalgorithm}
\end{figure}

\begin{figure}[t!]
  \centering
  \includegraphics[width=0.6\textwidth]{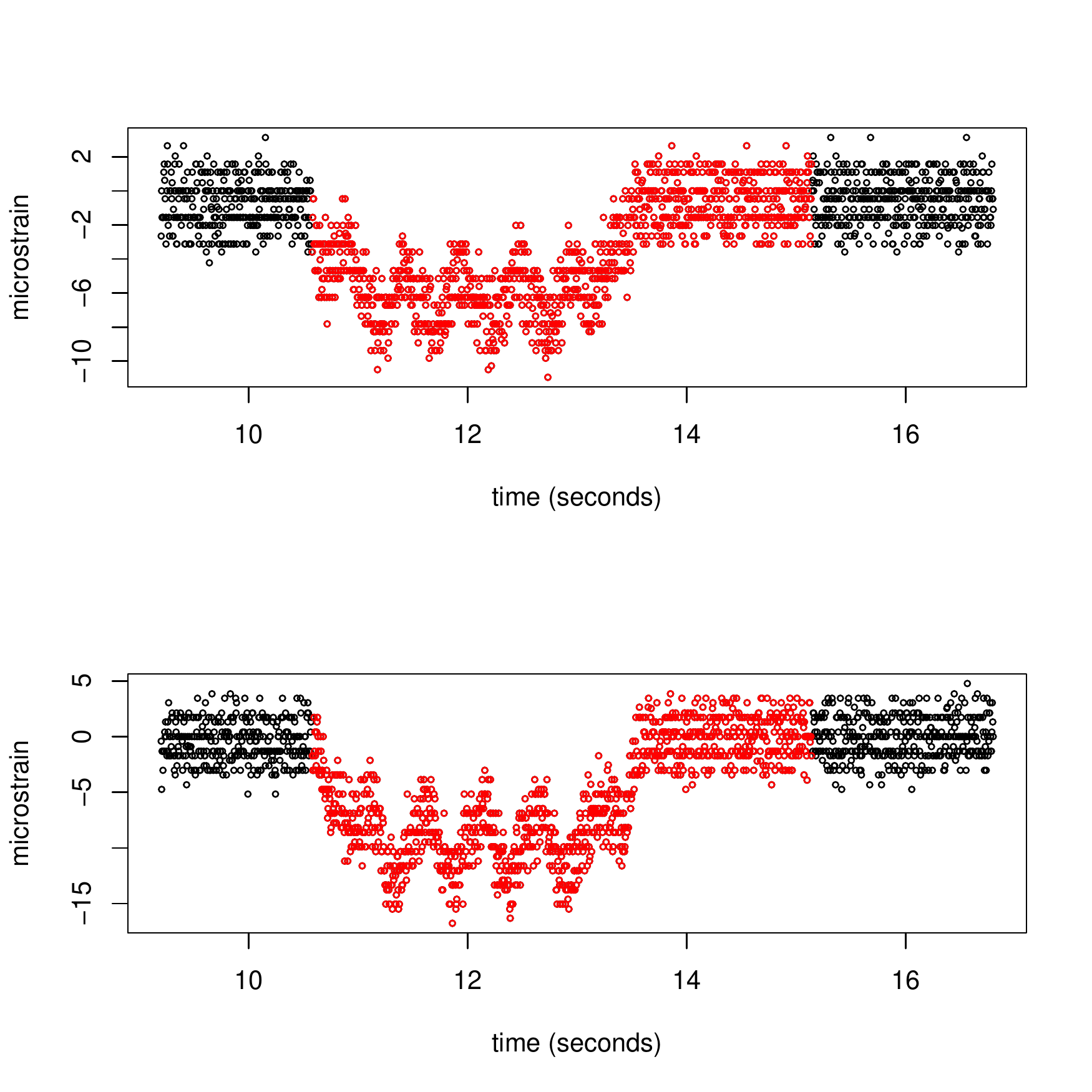}
  \caption{Magnified version of
    Figure \ref{fig:sensor_case_log_recoveryalgorithm}:  Data from sensor 1 (top panel) and sensor 20
    (bottom panel) during a restricted time period. The red data points show the estimated recovery period.}
  \label{fig:sensor_case_magnified_recoveryalgorithm}
\end{figure}

Figure \ref{fig:sensor_case_log_recoveryalgorithm} shows the estimated
recovery times for one of these events. In addition to $\log(X^2)$
over time, it shows the data streams from two sensors, 1 and
20. Sensor 1 is at the end of the girder where the train approaches
from. Sensor 20 is at the other end of the girder. Therefore the start
of the loading period of the train, and indeed the detected event
period, occurs in the data from sensor 1 slightly before the data from
sensor 20. This slight delay is noticeable in the magnified plots of
the data streams in Figure
\ref{fig:sensor_case_magnified_recoveryalgorithm}.

The algorithm successfully picks out the train passage event using the
statistical test. All 35 events were successfully detected by the
algorithm. Notice that the values of $X^2$ take slightly
longer to recover to their baseline distribution than the
loading period seen in the sensor data. 
This indicates that the recovery time is longer than the observable loading period. 
During the loading period on the bridge, the $p$-values computed for
each sensor become approximately zero, as one can see by the lack of
$X^2$ points for this period. We note that the signal obtained from
this test statistic during such an event is significantly more strong
in this work, where a non-parametric quantile model is used, than that
observed in \cite{Lau} where a linear model is used.

\subsection{Long and indefinite data streams}

\label{sec:longtimeseries}

In Sec.~\ref{sec:sensor-data}, the temporal variation of the sensor
strain data over long periods of time is discussed. The quantiles summaries used in Algorithm \ref{algo:algo1} assume that
the data does not exhibit temporal variations. Therefore, we seek
a baseline distribution of sensor data which does not exhibit temporal
variation, 
whilst maintaining the banding structure shown in
all of the sensor data. To achieve this, a moving-median detrending
scheme is now proposed. Later in the subsection this scheme is used
alongside Algorithm \ref{algo:algo1} to demonstrate the continuous
updating of a baseline sensor strain distribution, that is free from temporal variations, and the estimation of recovery times from any train
passage events. 

\begin{figure}
\centering
\begin{subfigure}{0.5\textwidth}
  \centering
  \includegraphics[width=0.8\textwidth]{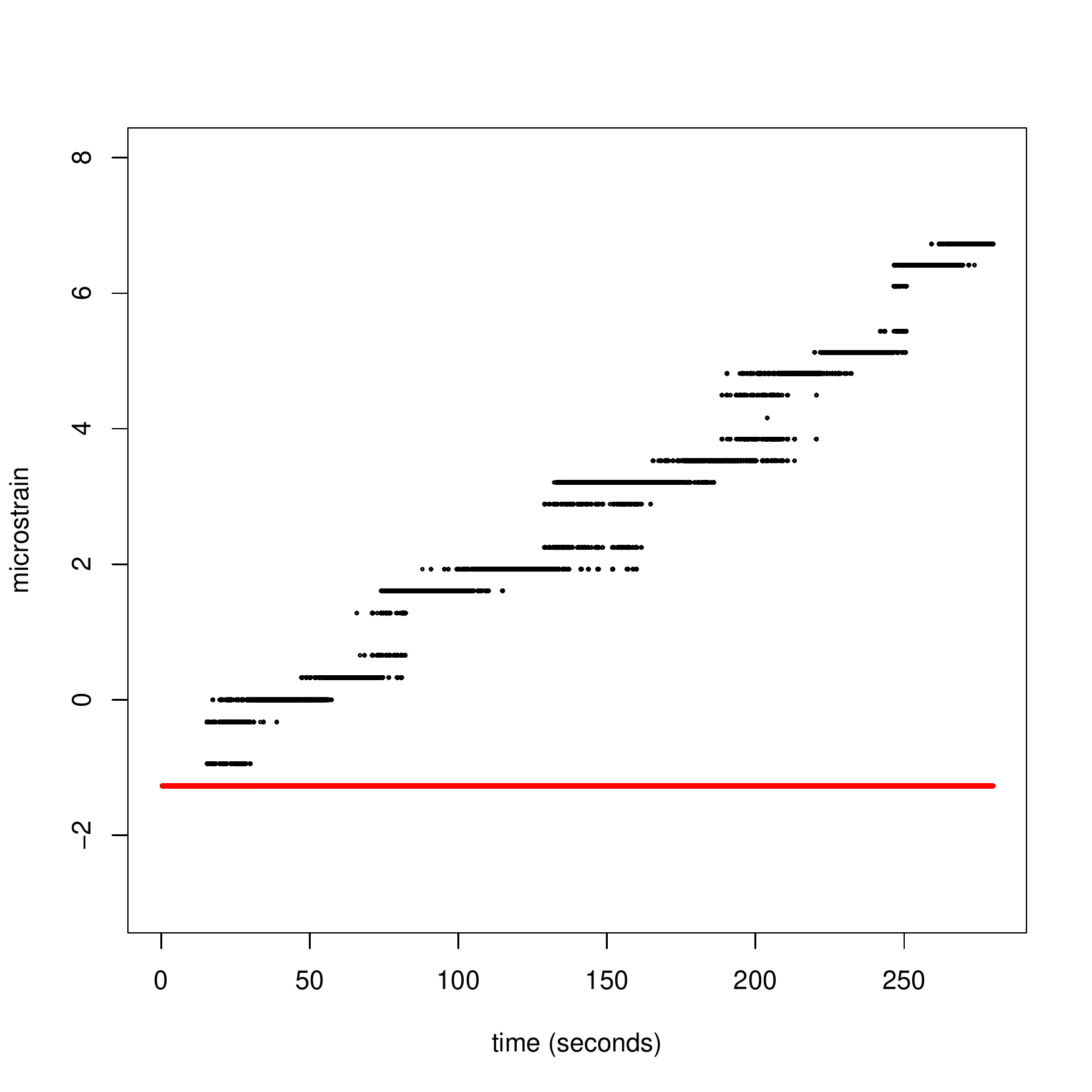}
  \caption{}
  \label{fig:detrended_long_sensor_dataa}
\end{subfigure}%
\begin{subfigure}{.5\textwidth}
\centering
  \includegraphics[width=0.8\textwidth]{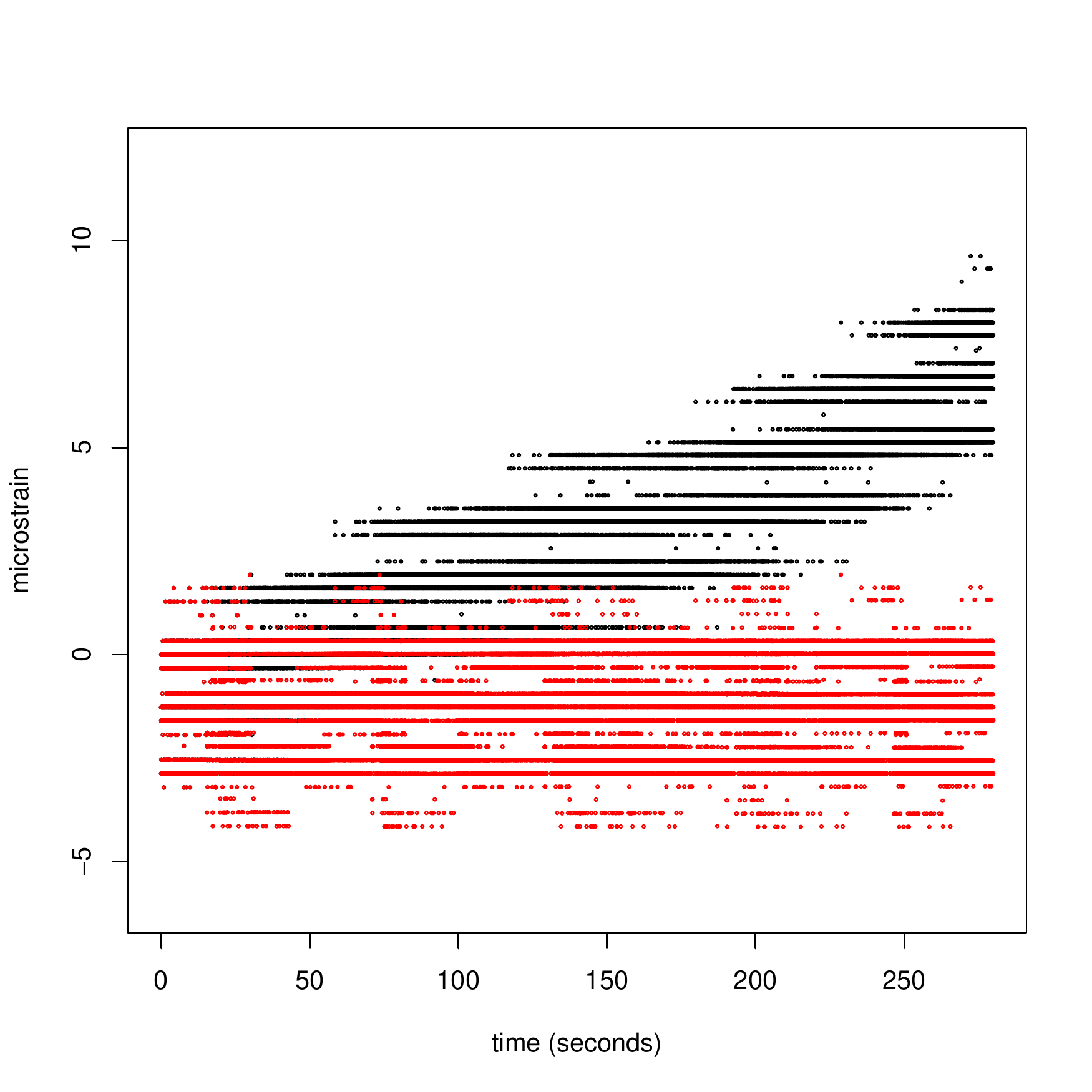}
  \caption{}
  \label{fig:detrended_long_sensor_datab}
\end{subfigure}
\caption{The values of $\mu_{\text{med}}(y^{(s)}_t)$ and $Q_{t}^{(s)}(0.5)$ (a) and the detrended sensor data, $\hat{y}^{(s)}_t$(b). The moving-median is shown by the black dots in (a) whilst the raw sensor data is shown by the black dots in (b). The value of $Q_{t}^{(s)}(0.5)$ is shown by the red dots in (a) whilst the detrended sensor data is shown by the red dots in (b).}
\end{figure}

Whilst implementing Algorithm \ref{algo:algo1} the detrended sensor data, $\hat{y}^{(s)}_{t}$, is now used when updating the quantile summary and computing $X^2$. This modified data is given by
\begin{equation}
\hat{y}^{(s)}_{t} =
\begin{cases}
y^{(s)}_t + \left(Q_{t}^{(s)}(0.5)-\mu_{\text{med}}(y^{(s)}_t,m)\right), & C=0\\
y^{(s)}_t + \left(Q_{t}^{(s)}(0.5)-\mu_{\text{med}}(y^{(s)}_{T_{S}},m)\right), & C=1,
\end{cases}
\end{equation}
where $\mu_{\text{med}}(y^{(s)}_t,m)$ is a moving-median of the sensor
data over the interval $[\max(t-m,1),t]$. The case when $C=1$ uses an
estimate for the median of the data immediately preceeding the event
in question (with start time $T_{S}$). This is because the sensor data
are not assumed to be from the baseline distribution during an
event. The window length $m$ is assumed small enough that the
space-memory required to store the stream values used in the
moving-median is negligible in comparison to that used by the quantile
summary. The modifications proposed above shift the temporally varying
sensor data by the difference between the moving-median and the median
of the baseline distribution. An estimate for the median of the
baseline distribution is stored in $Q_t^{(s)}$ (see
Sec.~\ref{sec:quantile_estimation}), and therefore no additional
computation is required. For the temporally varying sensor data shown in Figure \ref{fig:long_sensor_data_1}, where no train passage event occurs, the values of $Q_{t}^{(s)}(0.5)$ and $\mu_{\text{med}}(y^{(s)}_t,m)$ are shown in Figure \ref{fig:detrended_long_sensor_dataa} with $m=75$. The detrended sensor data corresponding to this data stream are shown in Figure \ref{fig:detrended_long_sensor_datab}; this baseline sensor distribution maintains the banded structure of the sensor data whilst excluding any temporal variability. 

\begin{figure}
\centering
\begin{subfigure}{0.5\textwidth}
  \centering
  \includegraphics[width=0.95\textwidth]{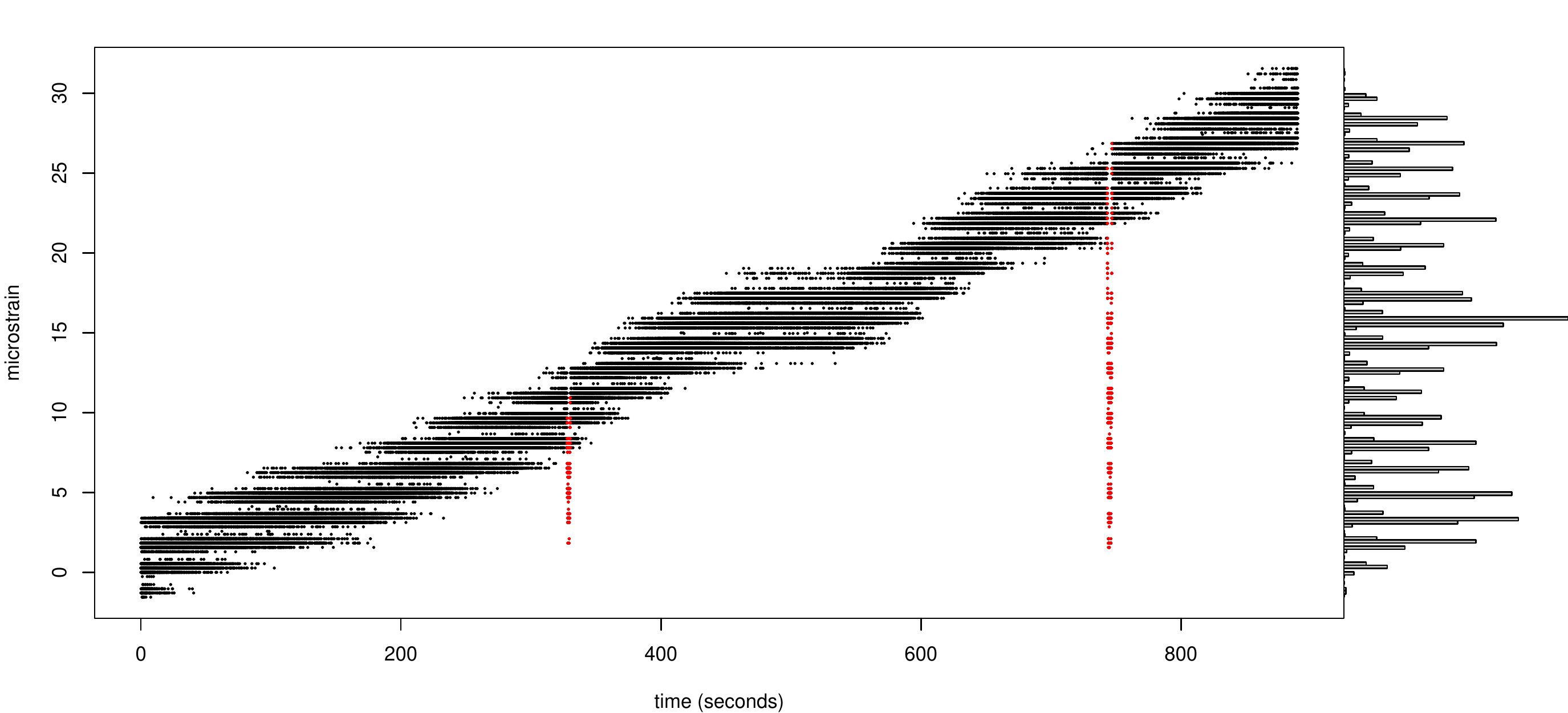}
  \caption{}
  \label{fig:detrended_data_long_time_seriesa}
\end{subfigure}%
\begin{subfigure}{.5\textwidth}
\centering
  \includegraphics[width=0.95\textwidth]{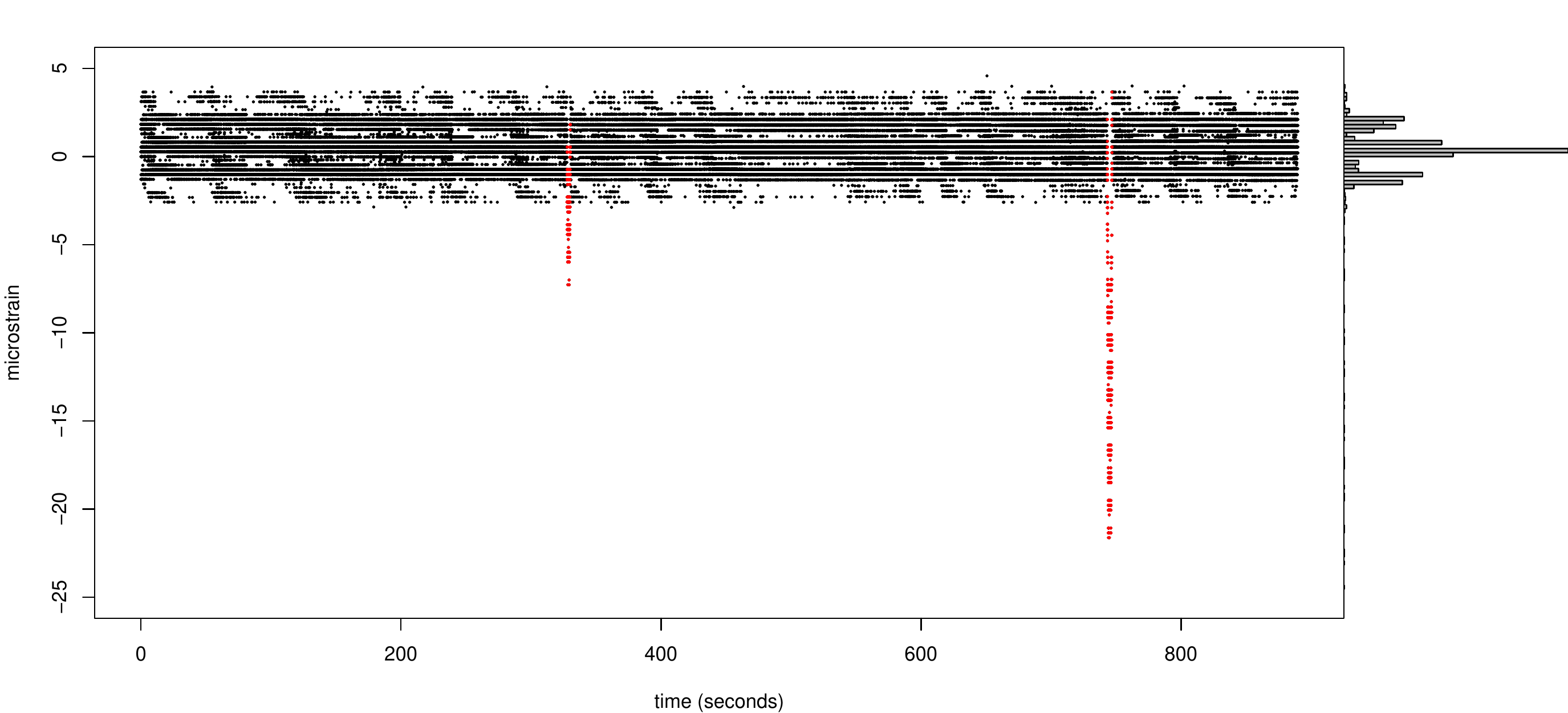}
  \caption{}
  \label{fig:detrended_data_long_time_seriesb}
\end{subfigure}
\caption{The raw data (a) and the detrended data (b) for a single FBG sensor over a long time interval exhibiting two train passage events. Histograms with 0.25-microstrain units wide bins for both sets of data are shown at the side of the plots. The detected train passage events are shown by the red points in both plots.}
\end{figure}

The detrended baseline sensor distribution can be utilised to model recovery times from train passage events in the same way as the distribution inferred by a short, temporally invariant data set (see Figure \ref{fig:sensor_case_log_recoveryalgorithm}). Figure \ref{fig:detrended_data_long_time_seriesa} shows the data from a single sensor for a temporally varying data set with 222,441 points. Two train passage events occur, and the algorithm detects both of them. Here detrending has been used to infer baseline sensor distributions. The corresponding detrended data are shown in Figure \ref{fig:detrended_data_long_time_seriesb}. Histograms with 0.25-microstrain units wide bins of both the raw and detrended data are also shown.

\begin{figure}[t!]
  \centering
  \includegraphics[width=0.7\textwidth]{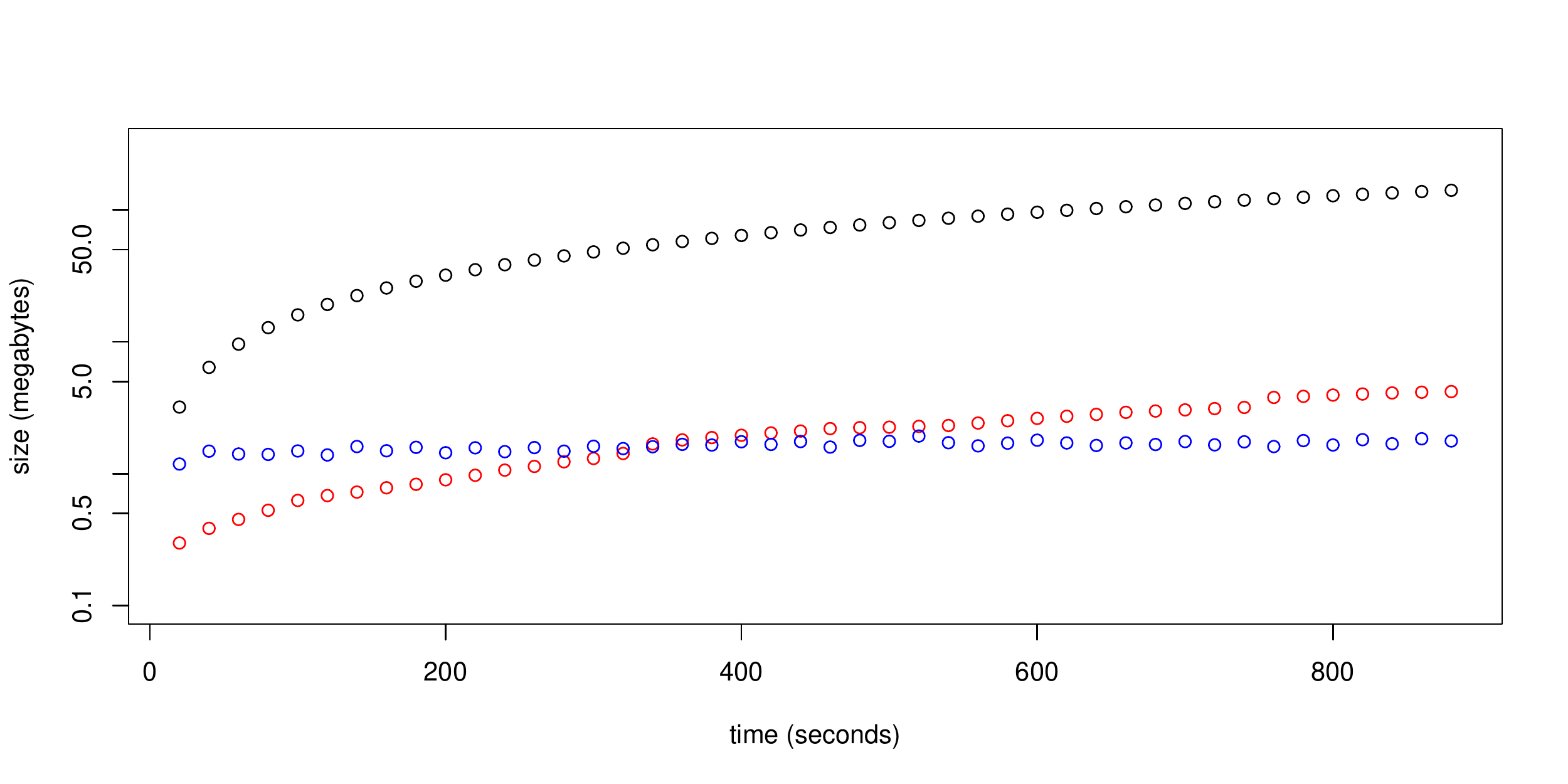}
  \caption{The combined size (in megabytes) of the detrended data sets from all sensors (black), the quantile summaries from all sensors (blue) and the frequency counts for the unique values of the detrended data sets from all sensors (red). Note that for long time series, maintaining quantile summaries for the baseline sensor distributions is space-memory efficient.}
  \label{fig:size_long_time_series}
\end{figure}

Figure \ref{fig:size_long_time_series} shows the size of the data
stored (in megabytes) for the quantile summaries, entire data set, and frequency counts of all of the unique
points in the data over time. Frequency counts can be used to
construct quantile estimates in discrete data streams
\cite{Shrivastava}. However the quantile summary is more space-memory
efficient than using the frequency count. The temporal variability of the raw sensor data means that the number of unique values in the data increases. This suggests that the methodology presented in this paper is more appropriate in the setting of indefinite data streams.


\section{Conclusion}

This paper has introduced a space-memory efficient, quantile-based
model for the baseline data from FBG sensors fitted to instrumented
railway bridges. A baseline distribution for data from each sensor is
iteratively updated in an efficient manner using an approximation to
the empirical quantile function. This method also removes the temporal
variation in the sensor data that are likely due to temperature
changes. A novel two-threshold statistical test is used to detect a change from the baseline
distribution to signal the start of a train passage event. Further,
the test detects a return to the baseline distribution to signal the
end of the event. Together, these event signals are used to estimate the
recovery time of the sensor network, and indirectly the bridge. Future
work will involve a more in-depth study of recovery times over longer
time periods to reason about long-term degradation. This study may include accounting for variables such as train length, train mass, train weight and data from temperature sensors.  


It is important to note that the statistical test utilised in the
algorithm presented in this paper assumes that all of the sensors are
independent of one another; this is not the case with the FBG sensors
instrumenting the bridge considered here. The dependent case of the
test can be implemented by Brown's Method \cite{Poole}. However, this
requires a recursive covariance computation and therefore is
computationally expensive for a large number of sensors. In order to
alleviate this issue, a future research direction could be to reduce
the dimensionality of the data before applying the methodology
presented in this paper. Similarly, another important extension of this research is the spatial modelling of the sensor network, rather than treating each sensor as individual components. The dense topology of this network can be utilised by spatial statistical models of bridge recovery times and the baseline distribution of other proxies for deterioration such as curvature and neutral axis position.

The model for the baseline sensor distribution is non-parametric and
features the banded and bounded structure that the data
exhibits. Retaining these features is not the case when the data are
modelled as, for example, a Normal linear model \cite{Lau}. The work
in \cite{Tee} uses a similar non-parametric approach to SHM where
regression models for a series of prescribed quantiles are utilised to
construct a damage detection algorithm. Our methods differs from this
work as we use an approximation of the empirical quantile function as
the model. We combined the Greenwald-Khanna algorithm to construct space-memory
efficient quantile function approximations for the baseline sensor
distributions and a detrending scheme to apply the methodology to long data streams. It is our hope that this direction of
research allows for the continuous recording of data from a sensor
network, without the concern of memory space.

\section*{Acknowledgement}

This work was supported by The Alan Turing Institute under the EPSRC grant EP/N510129/1 and the Turing-Lloyd's Register Foundation Programme for Data-Centric Engineering. The authors would also like to acknowledge EPSRC and Innovate UK (grant no. 920035) for funding this research through the Centre for Smart Infrastructure and Construction (CSIC) Innovation and Knowledge Centre. Research related to installation of the sensor system was carried out under EPSRC grant no. EP/N021614. Data related to this publication are available at the University of Cambridge data repository.

\bibliography{refs}
\end{document}